\documentclass[superscriptaddress]{elsarticle}
\usepackage{epsfig}
\usepackage{graphicx}
\usepackage{amsmath}
\usepackage{amssymb}
\usepackage{verbatim}   
\usepackage{color}      
\usepackage{subfigure}  

\newcommand{\beq}{\begin{equation}}
\newcommand{\eeq}{\end{equation}}

\begin{document}
\title{Dynamic Landscapes: A Model of Context and Contingency in Evolution}
\date{August 15, 2008, update \today} 

\author[label2]{David V. Foster}
\author[label3]{Mary M. Rorick}
\author[label4]{Tanja Gesell}
\author[label5]{Laura Feeney}
\author[label1]{Jacob G. Foster \corref{cor1}}
\cortext[cor1]{Please direct correspondence to jgfoster@uchicago.edu}
\address[label2]{Pluribus Systems, Durham, NC, USA}
\address[label3]{Department of Ecology and Evolution and Howard Hughes Medical Institute, University of Michigan, Ann Arbor, MI, USA}
\address[label4]{Max F. Perutz Laboratories, University of Vienna, Austria and School of Engineering and Applied Sciences, Harvard University, Cambridge, MA, USA}
\address[label5]{Swedish Institute of Computer Science and Uppsala University, Sweden}
 \address[label1]{Department of Sociology, University of Chicago, Chicago, IL, USA}


\begin{abstract}
The basic mechanics of evolution have been understood since Darwin. But debate continues over whether macroevolutionary phenomena are driven primary by the fitness structure of genotype space or by ecological interaction. In this paper we propose a simple, abstract model capturing some key features of fitness-landscape and ecological models of evolution. Our model describes evolutionary dynamics in a high-dimensional, structured genotype space with a significant role for interspecific interaction. We find some promising qualitative similarity with the empirical facts about macroevolution, including broadly distributed extinction sizes and realistic exploration of the genotype space. The abstraction of our model permits numerous interpretations and applications beyond macroevolution, including molecular evolution and technological innovation. 
\end{abstract}

\maketitle

\section{Introduction}
What drives evolution?  Since the days of Darwin, the prevailing explanation has emphasized heritable variation and selection. But while the mechanism of heredity and the importance of random mutation for generating variation have both been thoroughly explicated, the nature and causal agents of selection remain rather mysterious. As a result, we still struggle to explain those most striking events in the drama of life: the mass extinctions, adaptive radiations, and prosaic local speciations that generated the millions of species alive today and the hundredfold greater number that have become extinct over the course of evolutionary time \cite{GavRev}. 

Theories tackling these macroevolutionary questions fall into two broad categories. Each approach comes from a different intellectual tradition and has a distinct central metaphor. The first approach, derived from molecular genetics, focuses on the fitness landscape introduced by Sewall Wright \cite{Wright}. Selection cannot directly influence genotype frequencies; instead selection acts on the associated phenotypes. The fitness landscape metaphor therefore focuses on the map between genotype and fitness, as mediated by phenotype. More precisely, it constructs a map (the fitness landscape) that assigns a scalar fitness value to every point in genotype space. This fitness value guides the population-genetic process of mutation and selection occurring on the genotype space. Random mutants with higher fitness are favored by selection, on average, while those with low fitness are generally eliminated. 

The second approach, coming from ecology and dynamical systems theory, emphasizes the role of interactions in driving species to extinction. Here the genotype space is essentially ignored, and the fate of species in a given ecology (i.e., a pattern of interspecies interaction, often fixed) is determined by interacting population dynamics (e.g. the replicator dynamics of evolutionary game theory \cite{Sigmund}). Extinct species are usually replaced by a random extant species e.g. \cite{EvoSOC, SoleMan1}. As an example, abundant prey may drive the growth of a predator population, which in turn drives some third prey species to extinction. 

Both approaches capture important qualitative features of macroevolution. Given recent insight into the subtleties of the genotype-phenotype map, e.g. its many-to-one character, it seems clear that the interaction between genotype (the realm of heredity and variation) and phenotype (the realm of selection) is of central importance to phenomena like speciation, adaptive radiation (e.g. the Cambrian explosion \cite{Gould}), and punctuation in the rate of evolution \cite{EvoSOC, TopPos}. Likewise, variants of the ecological approach have established an important role for interspecies dynamics in generating extinction, up to and including mass extinctions that eliminate most biodiversity. In other words, both models capture important aspects of context (e.g. particular ecological configurations rendering some genotypes unfit) or contingency (e.g. particular mutational histories limiting or enhancing accessible genotypes) in evolution. In this paper we marry the stylized facts captured by the two approaches into a simple, unified model of evolutionary dynamics. Briefly, our model describes evolution on a richly structured fitness landscape, where context and contingency strongly determine the subsequent evolution of the simulated biosphere. Our model captures some qualitative features of evolution that have been observed empirically. We find a broad distributions of extinction sizes, and evidence that evolution takes place as an ``advancing front" through genotype space (see below).

In Section II we summarize the fitness landscape and ecological models discussed in the Introduction. In particular, we gather the stylized facts that guide the construction of our model. In Section III we describe the model and the numerical methods used in our simulations. Section IV reviews the results, and in Section V we discuss these results and offer our conclusions as well as an outline of future work.

\section{Background}
Our model of evolutionary dynamics builds on two important traditions in the mathematical modeling of evolution. In both cases, researchers have sought to translate invariant and universal features of the evolutionary process into mathematics. In reviewing these models, we seek a list of ``stylized facts" that can inform a parsimonious, realistic abstraction of the evolutionary process. 

\subsection{Fitness Landscapes}
Since their introduction by Sewall Wright \cite{Wright}, fitness landscapes have played a dominant role in evolutionary theory \cite{GavRev}. This dominance follows from their exceeding conceptual simplicity: the genetic code of the organism defines a space of potential genetic configurations or genotype space (denoted $\cal{G}$ henceforth). A fitness function $\Phi$(related to the likelihood of survival) is defined on the genotype space. This fitness function is a map $\Phi: \cal{G} \to $ $\mathbb{R}^+$ from the genotype to some scalar measure of fitness. A population of individuals is then defined over $\cal{G}$, and their population dynamics is influenced by $\Phi(\cal{G})$. 

In much of the literature following Wright's original suggestion, the fitness landscape is metaphorically described and mathematically modeled as a literal geographical landscape: these so called ``rugged landscapes" have many adaptive peaks, of various heights, separated by adaptive valleys \cite{GavRev}. But this simplification of Wright's picture, originally intended to act only as explanatory metaphor, has several major flaws when used as an explicit mathematical model for adaptive landscapes \cite{Pigliucci,Pig2}. First, low-dimensional models (of the sort suggested by direct interpretation of the geographic landscape metaphor) fail to generate speciation events with any reasonable probability. Selection pushes the population up adaptive peaks and away from adaptive valleys, so that crossing a valley to a new (perhaps better) peak is unlikely. Furthermore, shifting balance (wherein the population is subdivided so that stochastic shift across a valley is more likely and higher fitness types can then sweep the population) and founder effect speciation (wherein a small number of individuals found a new, population in which the probability of crossing a valley is enhanced) both fail to explain the observed fecundity of the biosphere \cite{GavRev}.

The second flaw also follows directly from the oversimplified "peaks and valleys" landscape picture. $\cal{G}$ is in fact enormously high dimensional, as most organisms have thousands of genes and millions or billions of base-pairs (hence dim $\cal{G}$ $\sim 10^6 - 10^9$) \cite{GavRev}. Fisher already observed that high dimensionality converts ``adaptive peaks" into saddle points, and makes a single peaked landscape, albeit in enormously high dimensions, more likely \cite{GavEnc}.  

Third, Kimura's claim that most evolutionary change is neutral (i.e. indifferent with respect to fitness \cite{Kimura}) seems to have been at least partially validated by extensive experimental evidence of neutrality. For example, the genotype-phenotype map for RNA and proteins is now known to be many-to-one (here the phenotype is the fold of the RNA or protein). This fact implies that many sequences are selectively neutral  \cite{TopPos}. Neutrality, if taken to its extreme, would lead to a totally flat fitness landscape rather than a rugged one, and selection would play no role whatsoever. 

An important compromise embracing ruggedness, high dimensions, and neutrality was suggested by John Maynard Smith. He noted that functional phenotypes must ``form a continuous network which can be traversed by unit mutational steps without passing through nonfunctional intermediaries" \cite{JMS}. The essence of this suggestion --- that genotype space is percolated by a network or networks of more-or-less equally fit genotypes, which nevertheless represent a small fraction of all possible genotypes --- forms the core of the neutral network or holey landscape approach pioneered theoretically by Gavrilets \cite{GavRev, GavEnc, Gav1, Gav2}. In this approach, selection plays a role: it defines the neutral network(s) and keeps populations from mutating into the ``holes" of the landscape (alternately, it drives rapid evolution out of the holes onto the ridges). Neutrality also plays its part, since most evolution takes place neutrally along the interweaving networks which define the adjacency and accessibility of various morphological types associated to the networks. Some advocates of the neutral network picture go so far as to claim that the structuring of the genotype space by these neutral networks plays a primary role in shaping the phenomena of speciation, adaptive radiation, and punctuated equilibrium \cite{TopPos}. Even adopting a moderate version of this view makes clear that the continuous, topologically trivial fitness landscapes that characterize the ``peaks and valleys" model misrepresent the actual structure of accessible genotype space.

The holey landscape picture is amply supported by both theoretical evidence (in which neutral networks seem to be an inevitable consequence of a surprising variety of model specifications) and by empirical evidence from studies of RNA and proteins \cite{GavRev, Gav1, Gav2, Born}. Later in the paper we shall present an extended interpretation of our model in terms of both proteins and RNA, in which specific evidence is reviewed. Theoretically, we follow the very simplest variant of the holey landscape, which Gavrilets calls Russian roulette: each genotype is assigned a fitness of $1$ with probability $p$ and $0$ with probability $1-p$ \cite{GavRev}. This variant assumes that small differences in fitness are insignificant compared to the difference between viable and inviable genotypes. Summarizing the stylized facts: we want

\begin{itemize}
\item selection to matter, while ignoring small differences in fitness;
\item very few genotypes to be fit;
\item the genotype space to be suitably high dimensional;
\item neutrality to play a substantial role;
\item neutral \emph{networks} to exist in the genotype space.
\end{itemize}

In Section III it will become clear that these facts lead naturally to several components of our model. 
As emphasized by \cite{Pig2}, however, one of the major shortcomings of the holey landscape model is its omission of interactions between genotypes, i.e. of ecology. 

\subsection{Ecological Models}
Fitness landscape models focus on genotype space and the way it is structured by selection, thus highlighting speciation. Ecological models focus on interspecific interaction, thus highlighting extinction \cite{EvoSOC, SoleMan1}. Evidence for extinction played an important role in overturning the static pre-Darwinian biology. More recently, the discovery of mass extinctions (most famously the KT extinction) raised questions about their cause. These events were originally explained by appealing to catastrophes. It now seems possible that ecological interactions alone can account for mass extinction. For example, while the asteroid impact that occurred around the time of the KT extinction may have caused it \cite{SoleMan1}, physics models of self-organized criticality suggest that these mass extinctions can also be caused by the same mechanism that operates at small scales, i.e. ecological interaction between species \cite{EvoSOC, SoleMan1}.

The Bak-Sneppen model is perhaps the most famous ecological model of evolution \cite{BakSnep}. Here $N$ species are arranged on a periodic lattice (a circle in the simplest case); neighbors on the lattice ``interact" ecologically. Each species $i$ is assigned a ``fitness" $B_i$, which represents a barrier to mutation (here, ``mutation" refers to the replacement of a species by a new, phenotypically distinct species;  one possible interpretation is the fixation of a novel phenotypic type within a population of individuals). The probability of mutation $p_i \sim e^{-B_i/T}$ depends exponentially on the size of the barrier $B_i$, with $T$ setting the timescale of mutations. This dependence on barrier size creates an exponential separation of timescales, so the species with the lowest barrier $B_{low}$ at a given time step is always assumed to mutate. This species and its neighbors are assigned new, random fitness values (the mutation in $B_{low}$ presumably alters the fitness landscape of its neighbors). The model self-organizes into a critical state characterized by scale-invariant power laws, with the exponents of the power laws sensitive to the dimension of the lattice (e.g. the circle) defining the interaction structure \cite{EvoSOC}. For example, if the lifetime of a species is defined as the time between two mutations at a given site on the lattice, then the distribution of lifetimes $t$ is $N(t) \sim t^{-\alpha}$ with $\alpha = 1.1$ for the 1-d circular lattice. The empirical distribution for fossil genera lifetimes has exponent $\alpha = 2$ \cite{EvoSOC}. Now define an avalanche as a series of mutations that are causally related (i.e. a change in species $i$'s barrier makes $j$'s the least fit, causing it to change, in turn causing one of \emph{its} neighbors to change, etc.). The size of an avalanche $s$ is thus the number of causally related mutations forming a given avalanche. The distribution of avalanche sizes also obeys a power law $N(s) \sim s^{-\tau}$ with $\tau = 1.1$ \cite{EvoSOC}. By contrast, the empirical $\tau \approx 2$ \cite{SoleMan1}, where $s$ here represents the size of an extinction event . 

Sol\'{e}, Manrubia and collaborators defined a somewhat more realistic ecological model of extinction and speciation, which has explicit population dynamics as well as some notion of inheritance (i.e., mutants retain many of the properties of the parent species) \cite{SoleMan1}. Their model is based on a connectivity matrix $\gamma_{ij}$ valued on the interval $(-1,1)$. These matrix elements represent the interactions between species, and are not strictly speaking ``food web" interactions but rather some generalized positive or negative influence of species $j$ on species $i$. Species viability is binary: $S_i = 0$ or $S_i = 1$ (extinct or extant, respectively). Using a step function $\Phi(z) = 1, z>0$, they define dynamics on this ecology \cite{SoleMan1}:

\begin{equation}
S_i(t+1) = \Phi\left( \sum_{j=1}^{n} \gamma_{ij}(t) S_j(t) \right)
\end{equation}
The model proceeds by: first, randomly varying the connectivity matrix $\gamma_{ij}$ (external driving); then implementing the population dynamics, which may render some species extinct; and finally replacing extinct species with $\epsilon$-varied mutants of an extant species randomly chosen to undergo adaptive radiation (i.e. the mutants copy the interaction pattern of their parents, subject to small variation). Note how adaptive radiation in this model differs from the Bak-Sneppen model. In the Bak-Sneppen model, the fitness of a mutant is not correlated with the parent species it is replacing, and the fitness of its interaction partners is randomized. In the Sol\'{e}-Manrubia model, the fitness of a mutant is uncorrelated with the species it replaces, but it \emph{is} correlated with the parent species, whose interaction pattern in copies with small variation. This ecological model reproduces the power law of extinctions with $\tau = 2.05 \pm 0.06$, consistent with fossil evidence \cite{SoleMan1}. These authors also track the role of positive and negative interactions in supporting and destabilizing the ecology, respectively. There are a number of extensions  and close relatives of this model \cite{SoleMan2, MayaMan, NewmanRob}.

The Tangled Nature model is another response to the Bak-Sneppen model \cite{Jensen} and also aims for greater biological realism. Like the Bak-Sneppen and Sol\'{e}-Manrubia models, it describes evolutionary dynamics in the context of complex ecological interdependencies, and shows that these coevolutionary interactions can account for the intermittency of evolution as revealed in the fossil record \cite{Hall, Christensen}. Unlike the Bak-Sneppen and Sol\'{e}-Manrubia models, which track the mean phenotype and ecological interactions of populations (i.e., species), the Tangled Nature model tracks individual genotypes within a highly multi-dimensional genotype space. In the Tangled Nature model, a high-dimensional genotype space (typically containing about ${10}^{6}$ distinct genotypes) is sparsely occupied by individuals (typically about 1000), and these individuals die and reproduce in each time step. Replication is associated with an error rate that sends new mutant individuals onto nearby sites within genotype space according to a per-site mutation rate (i.e., double, triple, etc. mutants are possible and increasingly likely as mutation rate increases). Individuals that occupy the same site within genotype space constitute a "species", and these individuals intra-specifically compete: replication rate is constrained by a constant carrying capacity at each site. Individuals occupying different sites (i.e., having distinct genotypes) interact through asymmetric ecological ``couplings" that positively or negatively influence the reproductive rates of individuals in a frequency-dependent manner. In this model, no mutations are inviable, and the rate at which individuals die is constant across sites and time, so the fitness landscape is only shaped by the effect of these intra- and inter-specific ecological interactions on the reproductive rate of individuals. While the ecological interactions between distinct genotypes are chosen randomly and hardwired (time-invariant), the interactions involving a given site are only ``activated" in proportion to the population sizes of the sites involved. Because this model does not explicitly consider lethal mutations, it emphasizes the gradual changes to genotype fitnesses (and thus the fitness landscape) that are due to changes in population size at the various currently occupied genotypes. 

The Tangled Nature model is thus able to go farther than the Bak-Sneppen model in showing that that micro-dynamics at the level of individuals (i.e., mutation in combination with ecological interactions between specific genotypes) can produce macro-dynamics at the level of species (i.e., the emergence of genotype clusters, mass extinctions, punctuated equilibrium, etc.). Specifically, the Tangled Nature model shows long periods of evolutionary stability followed by hectic periods of rapid change. It also displays a gradual increase in ``ecosystem stabilityÓ through time (the stable periods become longer, and the extinction rate decreases). There is a slight bias to positive ecological interactions in the occupation of genotype sites during evolutionary stasis \cite{Christensen,Hall}. 

The ``stylized facts" that we have drawn from the holey landscape paradigm differ from the framework of the Tangled Nature model on several accounts. First, we consider very low rates of mutational viability and very constrained movement by mutation through sequence space. Second, we consider a genotype space with many more dimensions than those typically simulated under the Tangled Nature model. Third, while the Tangled Nature model allows for direct fitness effects of mutations (e.g., a mutant is completely released from the carrying capacity-induced competition that constrains the reproductive rate of its parent), we approximate all viable mutations as selectively neutral (at least from the standpoint of first order selection). Fourth, while the Tangled Nature model includes two hierarchical levels of population structure governed by qualitatively different rules (i.e., individuals that co-occupy a given site versus individuals that occupy distinct sites), we focus only on interactions among distinct genotypes and thus do not consider explicit population dynamics at the level of a single site/genotype. Our framework shares this simplifying assumption with the Bak-Sneppen and Sole-Manrubia models. 

It is not clear that evolutionary dynamics within a genotype space of size $\sim {2}^{20}$ that has fitness differences determined by intra-specific competition and hardwired ecological interactions among genotypes should be similar to evolutionary dynamics within a larger genotype space (of size ${2}^{50}$ containing only very few viable genotypes, the identity of which changes as a function of hardwired ecological interactions. Another reason to believe that the predictions of the Tangled Nature model might differ from a model that explicitly considers lethal mutations is that the Tangled Nature model is a frequency-dependent quasispecies model, so the emergent phenomena it describes (speciation and periods of evolutionary quasi-stability) are dependent on the mutation rate, and a classic error threshold exists for this dependence \cite{Avogadro}. The presence of lethal mutations has been shown to substantially effect quasispecies dynamics, particularly with regard to the location and existence of an error threshold  \cite{Tejero, Takeuchi, Wilke}. All of the above suggests that the predictions of the Tangled Nature model might substantially differ from the predictions of any model that incorporates a substantial number of inviable genotypes. Because most real-world evolutionary dynamics occur on holey fitness landscapes where this condition is met, examining ecological interactions within the framework of a holey landscape is an important theoretical next-step.

The perspective provided by ecological models is essential to a complete mathematical understanding of the macroevolutionary process \cite{PerMaya, Perspect}. We take from this literature the following stylized facts:

\begin{itemize}
\item interactions are an essential component of the evolutionary process;
\item these interactions should be generalized away from ``food web" pictures;
\item interactions should be able to render genotypes more or less viable;
\item interacting models tend to exhibit self-organization
\end{itemize}

An important element that we will not capture in the current model is the role of external driving. In \cite{SoleMan1} this appears through random changes in $\gamma_{ij}$. In \cite{NewmanRob} the driving is explicitly environmental. For simplicity we exclude these effects. Now we are ready to synthesize the stylized facts of Section II into a simple model. 

\section{The Model}

With our model we aim to unify the fitness landscape and ecological approaches to evolutionary dynamics. We recover some qualitative features of the empirical data about macroevolution, including some properties of extinctions and the way genotype space is explored by the population. In construction of the model, we will make frequent reference to the stylized facts that are being implemented at each step. 

\subsection{The Hypercube in dimension n}
The most primitive modeling choice concerns the structure of the genotype space $\cal{G}$. Since we aim at simplicity, we restrict ourselves to a binary genetic alphabet with no further structure (e.g. diploidy). The standard model for $\cal{G}$ under these considerations is the hypercube of $n$ dimensions, e.g. \cite{Eigen}. 

\begin{figure}
  \begin{center}
  \epsfig{file= 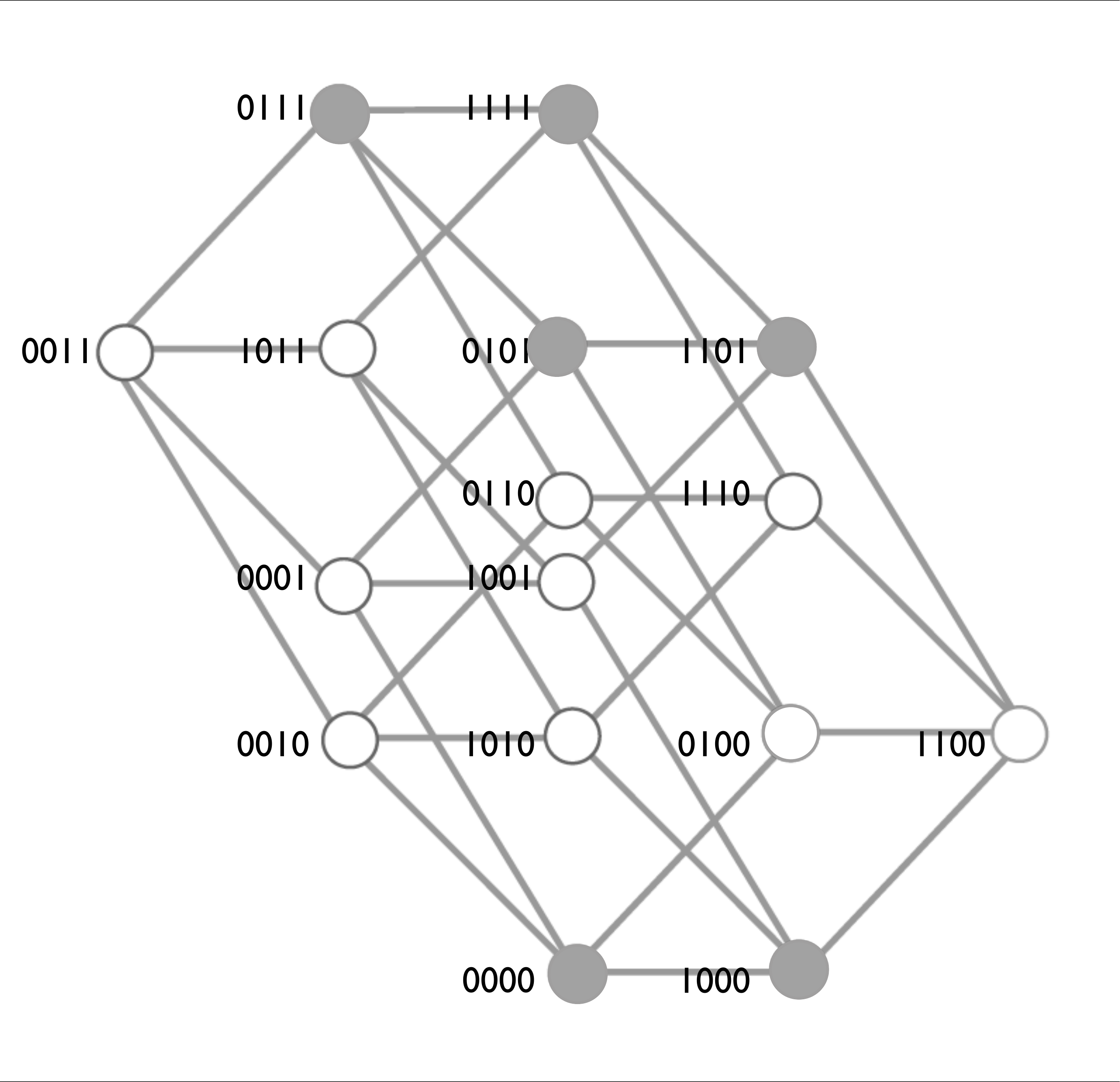, width=8.cm, angle=0}
  \caption{Hypercube of dimension 4. Note that each vertex is labeled with a string of $4$ $0$'s and $1$'s. Hence $1111$ has neighbors $0111$, $1011$, $1101$, and $1110$. The colored sites are elements of percolation clusters.}
  \label{hypercube_bare}
  \end{center}
\end{figure}

For a natural number $n$, we construct the graph $B^n$ ($n$-dimensional hypercube or $n$-cube) with vertices labeled by all $0,1$ sequences $(e_1,e_2,...,e_n)$ of length $n$; $e_i \in \{0,1\}$. Two $0,1$ sequences $s_i$ and $s_j$ are neighbors if they are of Hamming distance $H[s_i,s_j]=1$, i.e. if they differ  by one member of the sequence. For example, $0001100$ and $1001100$ are neighbors in the graph, and are hence connected by an edge. Note that each vertex has $n$ neighbors and $n$ is the ``coordination number" of the $n$-cube. The total volume (number of sites) of the $n$-cube is $2^n$. In Figure \ref{hypercube_bare} we show a labeled hypercube with coordination number $4$. 

This geometry models the genotype space of our universe of organisms. Because of the high level of abstraction, we can interpret the sequences variously: as a simple genetic alphabet; as hydrophobic or hydrophilic amino acids in a protein primary structure; or as some kind of binary morphological traits. 

The neighborhood relations indicate what sequences can be reached via a single point mutation. We do not permit insertions, deletions, or duplications, so the dimension $n$ remains fixed. We also forbid recombination (which can be viewed as a type of non-local mutation). Note that even for a relatively short sequence e.g. $n= 50$ the size of the space is enormous ($2^{50}$ possible genomes). Because our fitness function assignment is equivalent to percolation, in accordance with the terminology from this theory, we will refer to the vertices as ``sites" henceforth \cite{BakSnep}.

\subsection{Fitness function via Percolation}
Recall from the stylized facts that selection should matter, but that small differences in fitness should be irrelevant. Further, recall that few genotypes should be fit. Following Gavrilets, we thus independently assign fitness $1$ with probability $p$ and fitness $0$ and with probability $1-p$ to each site $s_i$, denoting this value $s_i(1, ,  )$ and $s_i(0, , )$, respectively (the notation will become clear as the exposition continues). 
Assigning a fitness of $0$ or $1$ is of course a drastic approximation to the actual complexity of the genotype-phenotype map; for example we have introduced no correlation structure on the fitness function (which one might expect \emph{a priori}). Correlations do not affect the qualitative behavior of the holey landscape models of Gavrilets \cite{Pig2}, and should at worst require a higher value of $p$ for a spanning cluster to appear (see below).

The fitness value assigned to each site is called the viability \cite{Gav1,Gav2}; the underlying assumption is that an overwhelming majority of sequences are non-viable, either due to developmental/folding errors (in the RNA, protein, and morphological cases), or due to lack of niche (in the morphological case). Hence the important distinction is not between the fitness levels of genotypes but rather between those viable genotypes (fitness $1$) and inviable ones (fitness $0$). 

\begin{figure}
  \begin{center}
  \epsfig{file=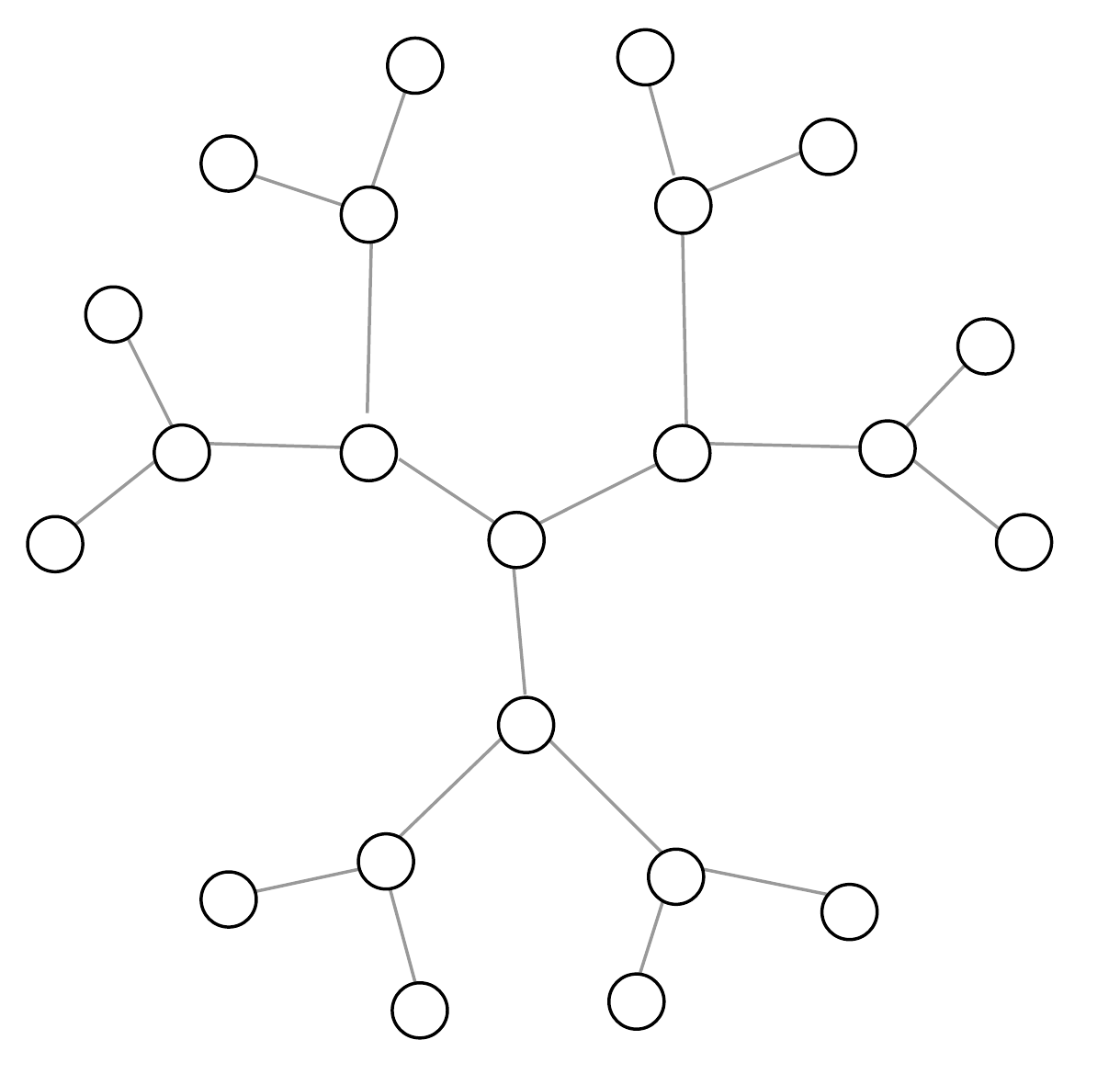, width=6.cm, angle=270}
  \caption{The Bethe lattice approximation. Here the coordination number is $n=3$. As $n \to \infty$ the approximation becomes exact.}
  \label{bethe}
  \end{center}
\end{figure}

This fitness function assignment is identical to a percolation process on the $n$-cube. If $H[s_i(1, , ),s_j(1, , )] = 1$ then $s_i$ and $s_j$ are neighbors in the same cluster of fit sequences. A cluster in our model (as in percolation) refers to a connected set of viable sites (fitness $= 1$), i.e., a set of viable sites such that there is a path from each site to every other site in the cluster that is a sequence of neighboring viable sites \footnote{Mathematically, $s_{k_1}$ and $s_{k_n}$ are in the same cluster if there exists a sequence of sites $ \{ s_{k_1},s_{k_2},s_{k_3}, ... , s_{k_n}\}$ such that $s_{k_{m}}$ and $s_{k_{m+1}}$ are both viable and $H[s_{k_m}(1, , ),s_{k_{m+1}}(1, , )] = 1$ for all $m$.}. A common tactic for dealing with very high dimensional spaces is to approximate the space by a tree with branching number equivalent to the coordination number. This is the so-called Bethe lattice approximation (Figure \ref{bethe}). The approximation becomes exact as $n \to \infty$. In the Bethe lattice approximation for ordinary lattices, one takes the limit of infinite lattice size (for lattices of sufficiently high coordination number) and hence expects to see finite \emph{size} corrections for any practical realization. In our case the thermodynamic limit of infinite size is equivalent to an infinite \emph{dimensional} limit, and there are finite dimension corrections \cite{PercHyp}. It is easy to calculate for the Bethe lattice the critical $p$ at which a ``spanning cluster" will appear, $p_c = 1/(n-1)$. At this $p$ any occupied site will have on average at least one occupied neighbor, so the cluster can persist indefinitely \cite{GavRev, Gav1, Molon}. We also note under the Bethe lattice approximation that \cite{Molon}:

\begin{itemize}
\item the average size of the cluster to which an occupied site belongs scales as $\chi(p) \sim (p_c - p)^{-1}$, $p \to p_c^-$;
\item the characteristic cluster size scales as $s_{\xi} \sim (p - p_c)^{-2}$, $p \to p_c^-$;
\item the cluster size density scales as $n(s,p) \sim s^{-\tau}exp(-s/s_{\xi})$, $\tau = 5/2$ close to $p_c$;
\item the ``surface area" or number of neighbors of a cluster of size $s$ is $t = 2+s(n-2)$.
\end{itemize}
The scaling of $n(s,p) \sim s^{-5/2}$ is confirmed in our simulations (see Results) and in studies of percolation on the hypercube in the context of spin glass relaxation \cite{PercHyp}. Note that as $n$ becomes large the fraction of occupied sites at the percolation threshold is correspondingly small ($1/(n-1)$), consequently satisfying our stylized fact that fit genotypes should be rare but that networks should nevertheless ``span" the genotype space. 

Here we have adopted the biological interpretation of a cluster as a ``neutral network". See Figure \ref{hypercube_bare} for an example with two percolation clusters. Point mutations from an extant viable genotype allow the occupation of neighboring viable genotypes. We will study the dynamics near the percolation threshold, as this is where the fitness function over the hypercube is interesting. For sub-critical percolation $p<p_c$ we expect many clusters, the largest clusters being of size $n$ \cite{Gav1}. We expect them to be relatively loop-free \cite{Gav1, PercHyp} for large $n$ (i.e. contain few pairs of sites linked by multiple paths within the cluster). Furthermore, for a randomly chosen site in $\cal{G}$, there should exist  some cluster passing within $n$ steps of that site \cite{Gav1}. For supercritical percolation, $p>p_c$, we expect the largest cluster to be of order $2^n/n$ for it to pass arbitrarily close to every site in the sequence space $\cal{G}$, and to be typically loopy \cite{Gav1}. 

In the subcritical case we can interpret a cluster as a ``species": a particular RNA or protein fold, in the molecular evolution context; a classical biological species, in the microevolutonary context; and a larger taxonomic unit, in the macroevolutonary context. Such interpretation is more difficult for the supercritical case without additional genetic complications as studied by Gavrilets \cite{Gav1,Gav2}. Here it makes more sense to view a cloud of neighboring occupied genotypes as a biological species \cite{Pig2}. Some empirical neutral networks can be thought of as a percolating cluster that is naturally partitioned into various species (fold) subclusters, which nevertheless percolate --- the subclusters are connected by one mutational step \cite{TopPos, Fontana, Babajide}. 

After setting up the neutral landscape in this fashion, we initialize the model by occupying a randomly chosen, viable site $s_0$. We will refer to an occupied site as an "extant genotype", with the caveat that it may refer to a sequence of base pairs, an amino acid sequence, or a particular collection of discrete morphological traits.

\subsection{Interactions}
So far the model construction has followed the fitness landscape tradition. Now we include complications coming from the ecological picture. Namely, we incorporate generalized interactions, which can render sites viable or inviable. 

We place a dotted directed edge between each ordered pair of distinct sites $(s_i,s_j), i \neq j$ with probability $q$  (note we independently try $(s_j,s_i))$. For each edge generated, the edge is $+$ with probability $0.5$ and $-$ with probability $0.5$. Thus we have a directed, signed Erd\"{o}s-R\'{e}nyi graph with expected edge number $\langle L \rangle = (2^n)(2^n - 1)q$ and with, on average, half $+$ and half $-$ edges. The mean degree is $\langle k \rangle = 
\langle L \rangle / (2^n)$, making no distinction on edge direction. 

If a site $s_i$ has incident edges (i.e. edges pointing towards it), and the sites from which these edges originate are occupied (call these ``activated" edges) we sum the total activated edge symbols with $+ = 1$ and $- = -1$. If the sum is positive, we write $s_i(e,+, )$, where $e \in \{0,1\}$ is the viability. If the sum is negative we write $s_i(e,-, )$. We interpret these symbols as follows. $s_i(0,+, )$ is {\it conditionally viable}, meaning that other occupied sites have created a niche for site $i$ (e.g. in the molecular RNA/protein interpretation, other molecular species facilitate a fold). $s_i(1,+, )$ is viable. $s_i(1,-, )$ is {\it conditionally inviable}, meaning that other occupied sites have eliminated the niche via predation, resource destruction, etc. $s_i(0,-, )$ is simply inviable. Note that $+,-$ is subject to change as the sources of other incident edges become occupied. Thus a viable site can become viable, then inviable, then viable again depending on the history of the system.

\subsection{Dynamics}
\begin{figure*}
  \begin{center}
  \epsfig{file=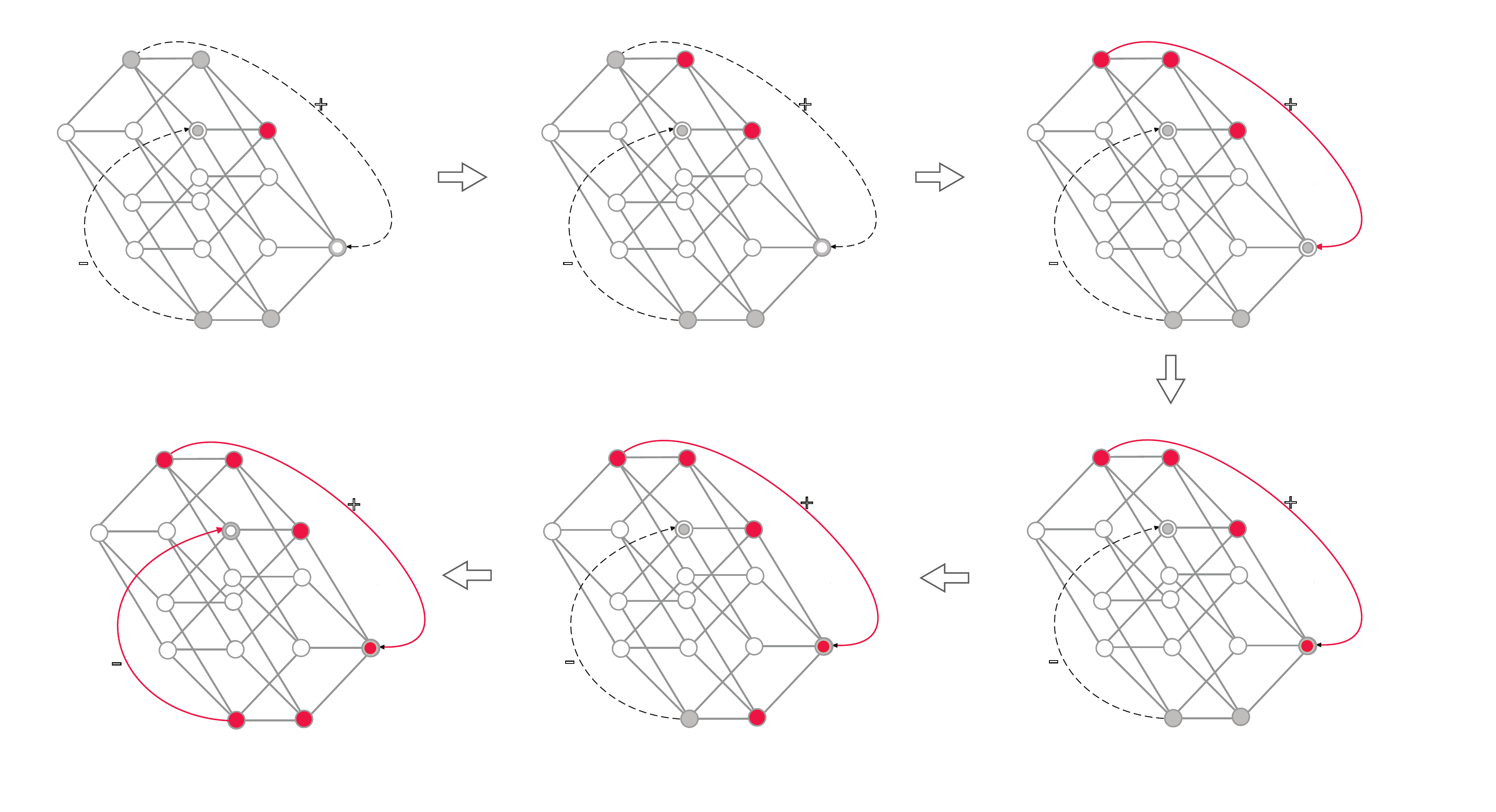, width=16.cm, angle=0}
  \caption{Model Dynamics for $n=4$. Vertices: grey = viable sites, grey with white boundary = conditionally viable sites, white with grey boundary = conditionally inviable, red = occupied sites.  Note the interactions rendering sites viable (thus connecting the upper and lower clusters) and inviable (eliminating a member of the upper cluster). Dotted lines: interactions increasing the fitness of a site ($+$) or decreasing it ($-$). A sequence of mutations creates a path through the hypercube, while interactions create a path between the upper and lower clusters (third step, c) or eliminate a member of the upper cluster (sixth step, f).}
   \label{animation}
  \end{center}
\end{figure*}

We have put extensive structure on our genotype space $\cal{G}$. In Figure \ref{animation} this corresponds to the grey vertices and dashed edges. We can now describe dynamics --- an important stylized fact of the ecological approach. Start the system by occupying some site $s_0(1, , )$. Notationally, the occupation of sequence $s_0$ can be written $s_0(1, ,1)$. This means that $s_0$ is viable, it has no interactions (because no other sites are occupied) and it is occupied. At every time step, we choose a random occupied site, i.e. extant genotype, pick one of its $n$ neighbors at random and send a mutant to that sequence, keeping the original site occupied. Note: 

\begin{itemize}
\item for realistic (small) mutation rates, the probability of two simultaneous mutations is negligible (thus justifying the choice of one genotype to mutate per time step);
\item having a mutant at every time step (as opposed to having some mutation probability) is just a matter of setting the timescale.
\item $p$ is very small so most mutations will fail;  
\end{itemize}
If the neighbor $s_i$ is viable or conditionally viable, i.e. $s_i = s_i(1,+,0)$ or $s_i(0,+,0)$ then the step is successful (the last element of the triple can be $1$ but then no new site is occupied). We update in this case

\begin{itemize}
\item $s_i(1,+,0) \to s_i(1,+,1) $ 
\item $s_i(0,+,0) \to s_i(0,+,1).$ 
\end{itemize}

Otherwise the mutation fails. Note that we allow ``back" mutations implicitly. In the second step of Figure \ref{animation} another site has been occupied by mutation from the initial site. 

At every occupation, the arrows originating from the newly occupied site become active and influence the sites to which they point. In the third step of Figure \ref{animation} the occupation of yet another site has rendered a site conditionally-viable (red $+$ edge pointing to a grey site with white boundary). Note that interaction can cause extinction as well as the creation of conditionally viable sites. In the sixth step of Figure \ref{animation} a sequence of mutations has rendered a site conditionally inviable (red $-$ edge pointing to a white site with grey boundary). We iterate these steps until no further successful mutation is possible. In principle, interactions will create ``bridges" across gaps between neutral networks, allowing major ``speciation" events to occur, as shown in the Figure. In other words, interactions (specifically, \emph{positive} interactions) can allow neutral networks to span the genotype space even in the subcritical case, by creating bridges out of an initial cluster into other clusters. \emph{Negative} interactions induce population dynamics in the model; specifically, they allow species to go extinct and sites to vary in their viability as a function of time.

\subsection{Numerics}
Because of the small value of $p$ most mutations will be failures. This follows from the stylized fact that few genotypes are viable. If we were to simulate the dynamics directly, as described in the text above, most computer time would be wasted on unsuccessful jumps to inviable sites. Consequently, we use an event-driven Gillespie scheme to select the next successful mutation and randomly generate an appropriate waiting time \cite{Gilles}. 

Specifically, every potential mutation from an occupied site $s_i$ to a viable but unoccupied site $s_j$ is assigned an equal propensity $pr(i,j) = 1$. All such ordered pairs have an equal probability of being selected for implementation. After a pair is selected, we generate the amount of simulation time $t(i,j)$ that passes before the event occurs. Let $X$ be a random number chosen from the uniform distribution over $[0,1)$  and $T = \sum pr(i,j)$, i.e. the number of potential viable mutation steps. Then the waiting time $t(i,j) = \frac{1}{T}\log \left( \frac{1}{X} \right)$ \cite{Gilles}. This approach conserves computing time by assuring that every simulation step involves a successful mutation. 

\section{Results}
Our model is parameterized by the dimension of the genotype space, $n$; the percolation probability, $p$; and the interaction probability $q$. It is obviously desirable to study $\cal{G}$ with as high dimension as possible. With available computational resources we were able to simulate $n=50$. Each $n=50$ simulation lasted approximately $70,000$ simulation steps, the limiting factor being memory usage. 

\begin{figure}
  \begin{center}
  \epsfig{file=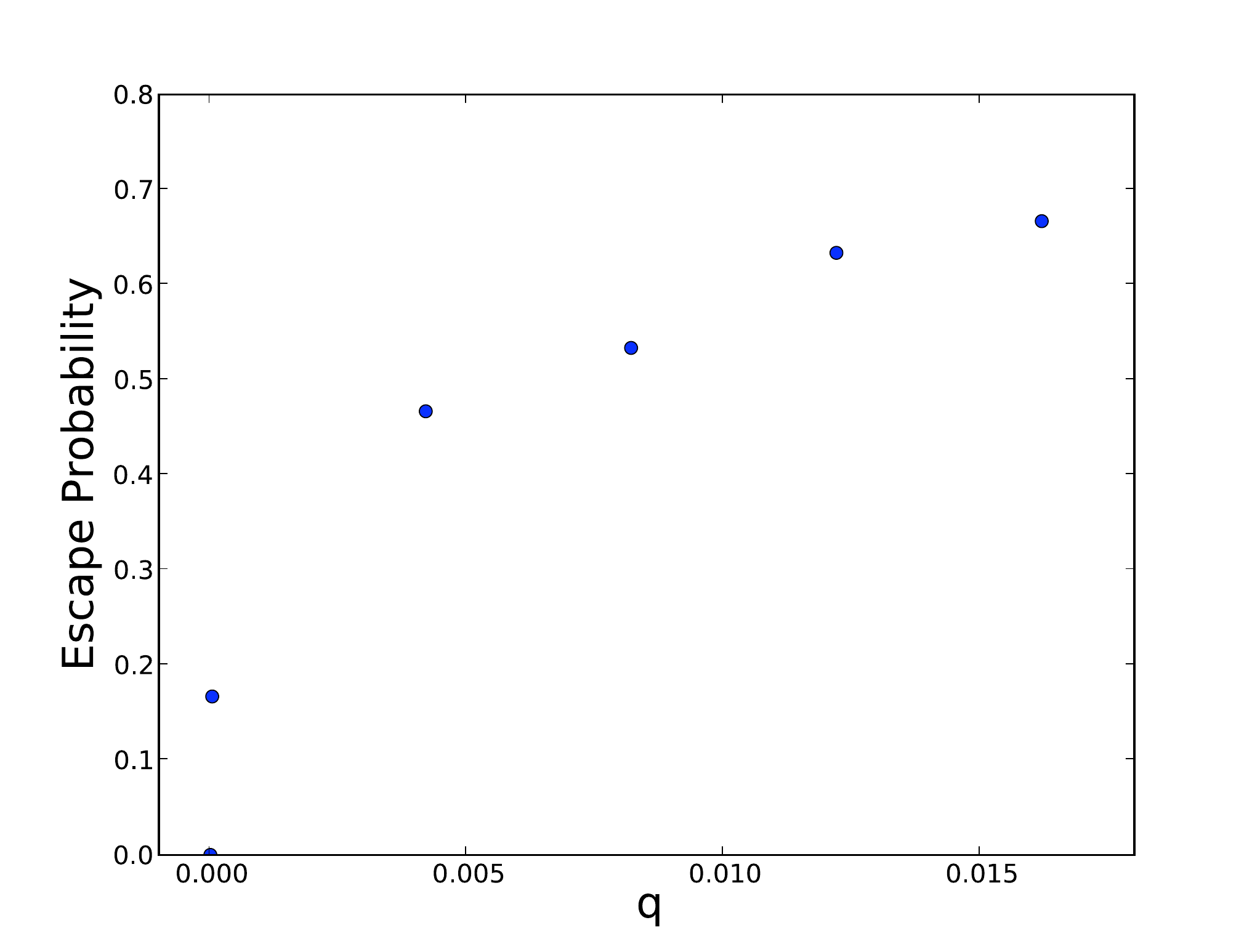, width=9cm, angle=0}
  \caption{In this plot we show the escape probability as a function of the interaction parameter $q$, which sets the likelihood for any ordered pair of sites to have an interaction. $n = 50$ and $p = 0.02037$, i.e. $p$ is slightly sub-critical. The first non-zero point on the plot is very close to $q = 2/n^2$, at which value most simulations were performed.}
  \label{escape}
  \end{center}
\end{figure}

We selected values of $p$ close to the critical value $p_c = 1/(n-1) = 1/49 \approx 0.02041$.  The choice of just sub- or super-critical $p$ appeared to have little impact on model outcome for fixed values of $q$. The selection of $q$ was slightly more difficult. For subcritical $p$ we scanned the parameter space of $q$ near $q = 1/n^2$, evaluating the probability that a simulation escapes the initial cluster, i.e. that interactions create a bridge from the initial cluster to a different cluster. Because we do not construct the entire genotype space beforehand in our simulations -- as this would be too expensive computationally --  we defined "escape" through an indicator: the successful occupation of at least $n^2$ sites. This number of sites is much larger than the size of any initial cluster in the subcritical case; occupying $n^2$ sites therefore indicates that the simulation has escaped the initial cluster with overwhelming probability. This ``escape probability" as a function of $q$ is plotted in Figure \ref{escape}. We find that indeed near $q = 1/n^2 = .0004$ the escape probability is non-zero. Most simulations were conducted for $q = 2/n^2$, where we expect between $10\%$ and $20\%$ of simulations to ``escape" the initial cluster.

It is easy to see why simulations tend to escape the initial cluster when $q \sim 1/n^2$. For $p$ close to $p_c$, the largest clusters will be of size $n$. The ``surface" of a large cluster contains approximately $n^2$ sites. For any viable site currently in the cluster, the probability that it will interact with at least one site on the cluster's surface is close to 1. If there are $n$ viable sites in the cluster and half of the interactions are $+$, then interactions will make an additional $n/2$ sites in the cluster surface conditionally viable (if $q = 2/n^2$ then $n$ sites will be added to the cluster surface). These new sites increase the surface area of the cluster, permitting further growth, on average. (Note that the $n$ newly viable sites on the cluster surface also interact with sites currently in the cluster; when they all become occupied, they render $1$ site within the cluster inviable, on average, for $q = 2/n^2$.  We ignore this, as well as interactions ``within" the cluster, in this heuristic argument). 

\begin{figure}
  \begin{center}
  \epsfig{file=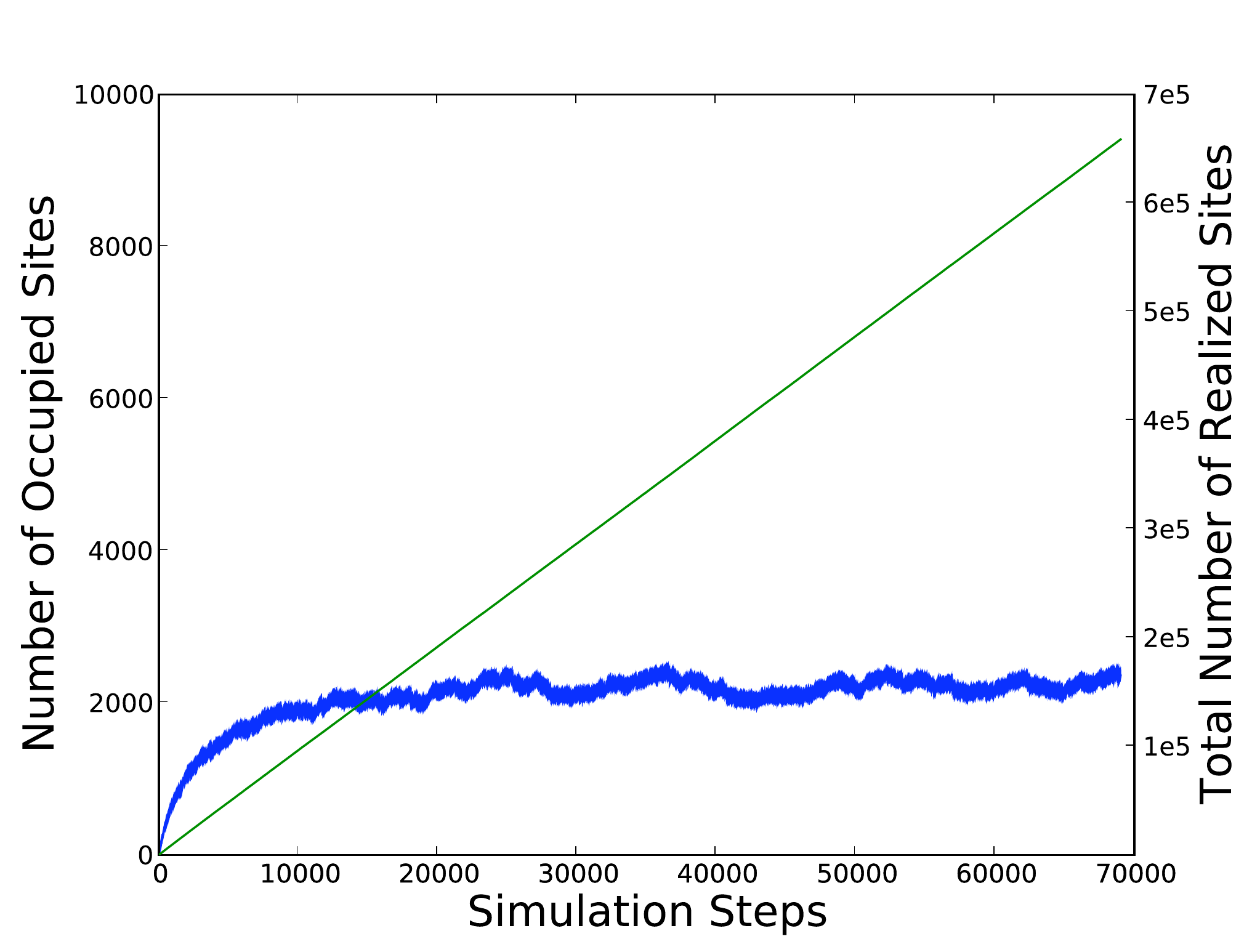, width=9cm, angle=0}
  \caption{The lower, blue line (left axis) shows the number of occupied sites (extant genotypes) as a function of simulation steps. Note that after an initial period of growth the number of occupied sites fluctuates around $2000$ at any given simulation step. The upper, green line (right axis) shows the number of realized sites, i.e. sites ever occupied in the course of the simulation, as a function of simulation steps. This quantity increases almost linearly with the number of simulation steps, indicating that evolution is continually exploring new parts of genotype space. Model parameters are $n=50$, $p = .02037$ (slightly subcritical) and $q = .0008 = 2/n^2$.}
  \label{Realized}
  \end{center}
\end{figure}

How are sites occupied over the course of a simulation run?  It is conceivable, for example, that most activity occurs ``near" the originating site, with sites continually discovered, rendered inviable, and then reoccupied when they become viable again. Keeping track of the number of realized sites, (i.e. sites that are \emph{ever} occupied in the course of a simulation) and the number of extant sites (i.e. sites occupied at a particular simulation step) reveals a very different picture, consistent with the heuristic picture of cluster growth described above. In Figure \ref{Realized} we see that the total number of realized sites (upper green line) increases almost linearly with the number of simulation steps. Hence at almost every simulation step the unfolding biosphere is discovering new sites or genotypes, rather than revisiting occupied territory. The fact that the number of extant genotypes (lower blue line) fluctuates around $2000$ sites over the course of the simulation indicates that there is an advancing front of activity, which explores genotype space while eliminating older genotypes as it advances. This behavior is typical of simulations that escape the initial cluster. Note that the steady state population of $2000$ sites is roughly the same size as $n^2 = 2500$. For $q=2/n^2$, any new site added will, with high probability, have one positive and one negative interaction with an existing site; thus it is unsurprising that the steady state is roughly $n^2$ sites, i.e. for every site added another site is rendered inviable.

Rather surprisingly, this picture of an advancing front is consistent with the morphological diversification of blastozoans discussed by Gavrilets \cite{GavMorph}. In this analysis the occupied sites are scored on a number of characters. Gavrilets then measures the average morphological disparity within extant genotypes, the taxonomic diversity, and the average morphological distance from the founder genotype. The picture described by Gavrilets of a ``compact group" moving away from the founder genotype in morphospace is extremely to what we observe in our simulation results. We could test the similarity of the two pictures by explicitly calculating at each simulation step the average Hamming distance from the founding site (among extant, occupied sites) and the average pairwise Hamming distance between extant genotypes. Gavrilets finds for the fossil data that the former increases while the latter grows initially but then shrinks and stabilizes at a small value. We will measure these quantities in future work; on the basis of the heuristic argument above, we predict behavior very similar to the fossil data.

We now turn to the role of interactions in facilitating the initial exploration of genotype space and subsequent stability at roughly $2000$ extant genotypes. Consider all pairs of viable sites $(i,j)$. Let the number of such pairs at time $T$ be $N(T)$. Define a time dependent quantity $f_T(i,j)$ taking the value $-1$ if the interaction is negative at time $T$ and $+1$ if the interaction is positive (recall that when set up, roughly half of the edges are positive and half negative; but we restrict here to edges between viable sites). Then the interaction bias between all pairs of viable sites $(i,j)$: 

\begin{equation}
I_{av}(T) = \frac{1}{N(T)}  \sum_{  i,j } f_{T}(i,j)
\end{equation}
provides a measure of the overall bias of interaction towards viability. If $I_{av}(T) = 0$ then positive and negative interactions between pairs of viable sites are perfectly balanced, and indeed absent any dynamics this is what we should expect, on average. 

\begin{figure}
  \begin{center}
  \epsfig{file=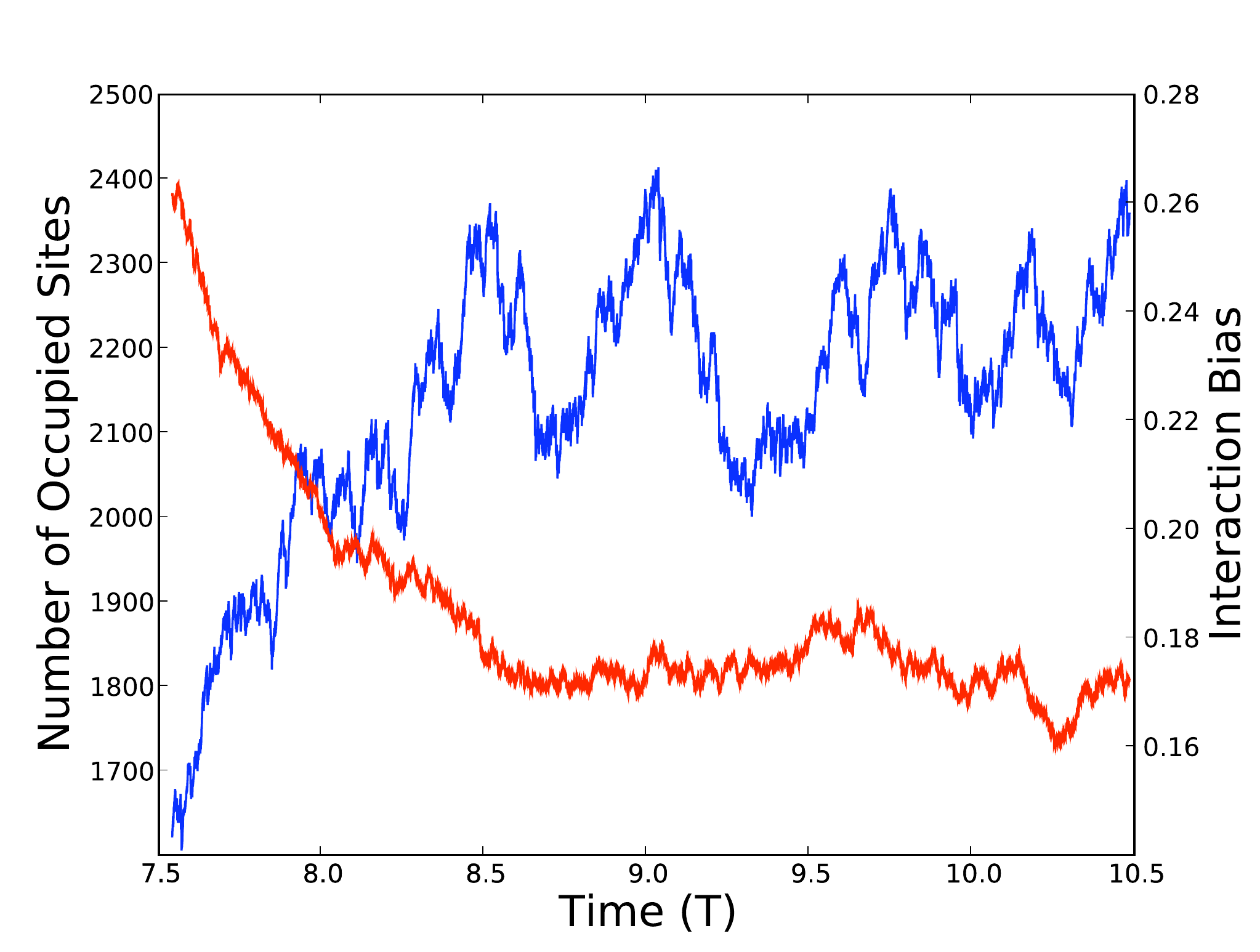, width=9.cm, angle=0}
  \caption{The relationship between number of extant genotypes and interaction bias. The blue line (upper after $T = 8.0$, left axis) shows the number of occupied sites (extant genotypes) as a function of simulation time $T$. The red line (lower, right axis) shows the interaction bias. Note that the interaction bias remains relatively constant as the simulation settles into the apparent steady state. Model parameters are $n=50$, $p = .02037$ (slightly subcritical) and $q = .0008 = 2/n^2$.}
  \label{Bias}
  \end{center}
\end{figure}

In Figure \ref{Bias} we see a typical time series (in fact the same time series from Figure \ref{Realized}) against which we also plot the interaction bias as a function of simulation time $T$ (rather than simulation steps). We have focused on the ``interesting region" after the simulation has grown towards its typical number of occupied sites. The period of growth is associated with a strong positive interaction bias, while the stabilization of the simulation into a steady state fluctuating around $2000$ occupied sites correlates with a decrease in interaction bias towards a small but nevertheless positive value. This behavior indicates that the steady state is ``supported" by a bias towards positive interactions. There are simultaneously enough interactions and enough occupied sites that a substantial number of negative interactions can take place between \emph{viable} sites without rendering those sites extinct. There seems to be some correlation between increases in the interaction bias and increases in the number of extant genotypes, and likewise between decreases in the bias and periods of sequential extinction, though this needs to be quantified. The apparent correlation is somewhat similar to the relationship between interaction entropy (similar to our interaction bias) and extinction measured in \cite{SoleMan1}. 

\begin{figure}
  \begin{center}
  \epsfig{file=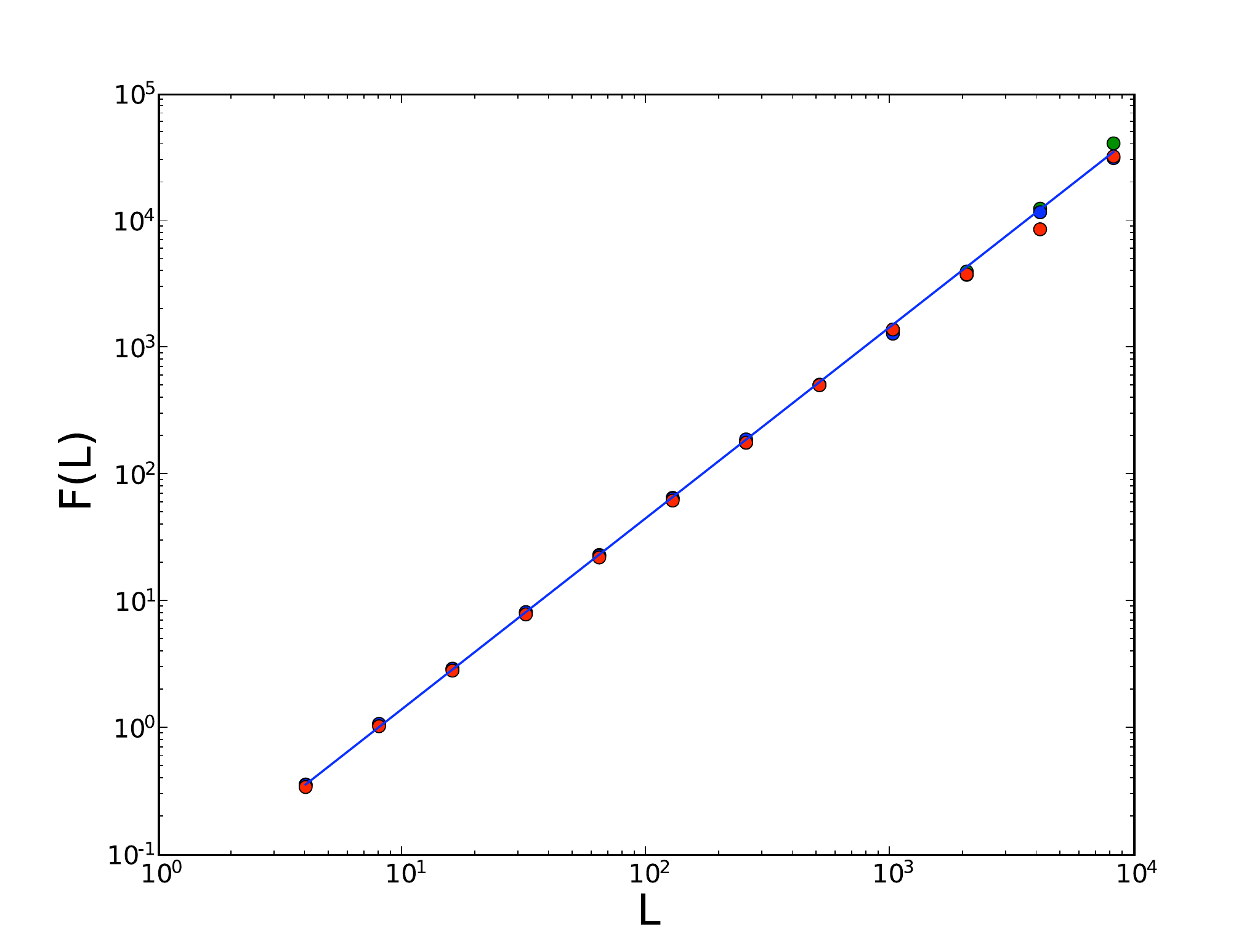, width=9.cm, angle=0}
  \caption{The detrended fluctuation analysis for three typical simulation runs. Red dots are for a genotype space of dimension $n=30$, blue dots are $n=40$, and green dots are $n=50$. $p$ is slightly subcritical, $p = 1/(n-.9)$ and $q= 2/n^2$ for the appropriate values of $n$. Note the logarithmic axes; all three plots are well fit by a power law with $\alpha \approx 3/2$.}
  \label{DFA}
  \end{center}
\end{figure}

In order to understand the behavior of the time series, we performed a detrended fluctuation analysis \cite{DFA}. We briefly summarize this technique. The data are divided into windows of length $L$. Within each window, we calculate a best-fit trend, in this case a polynomial of degree $2$. Then all fluctuations around the trend are summed for each window, normalized by the length of the window, and summed in turn, giving the total fluctuation $F(L)$. Plotting $F(L)$ against $L$ gives a power-law fit for self-similar time series, $F(L) = L^{\alpha}$. For our data we estimate $\alpha = 1.5065$ for simulations on a hypercube of dimension $n = 50$, see Figure \ref{DFA}. The fit of $\alpha \approx 3/2$ works well for $n=40$ and $n=30$ as shown in the Figure. $\alpha = 3/2$ indicates that the correlations in the time series are similar to those observed in Brownian motion. Generally, the time series exhibits fractal scaling insensitive to the dimension of the underlying genotype space. Comparison with detrended fluctuation analysis of taxonomic diversity data such as that from \cite{GavMorph} will be a fruitful direction for future work and a direct test of the model. 

\begin{figure}
  \begin{center}
  \epsfig{file=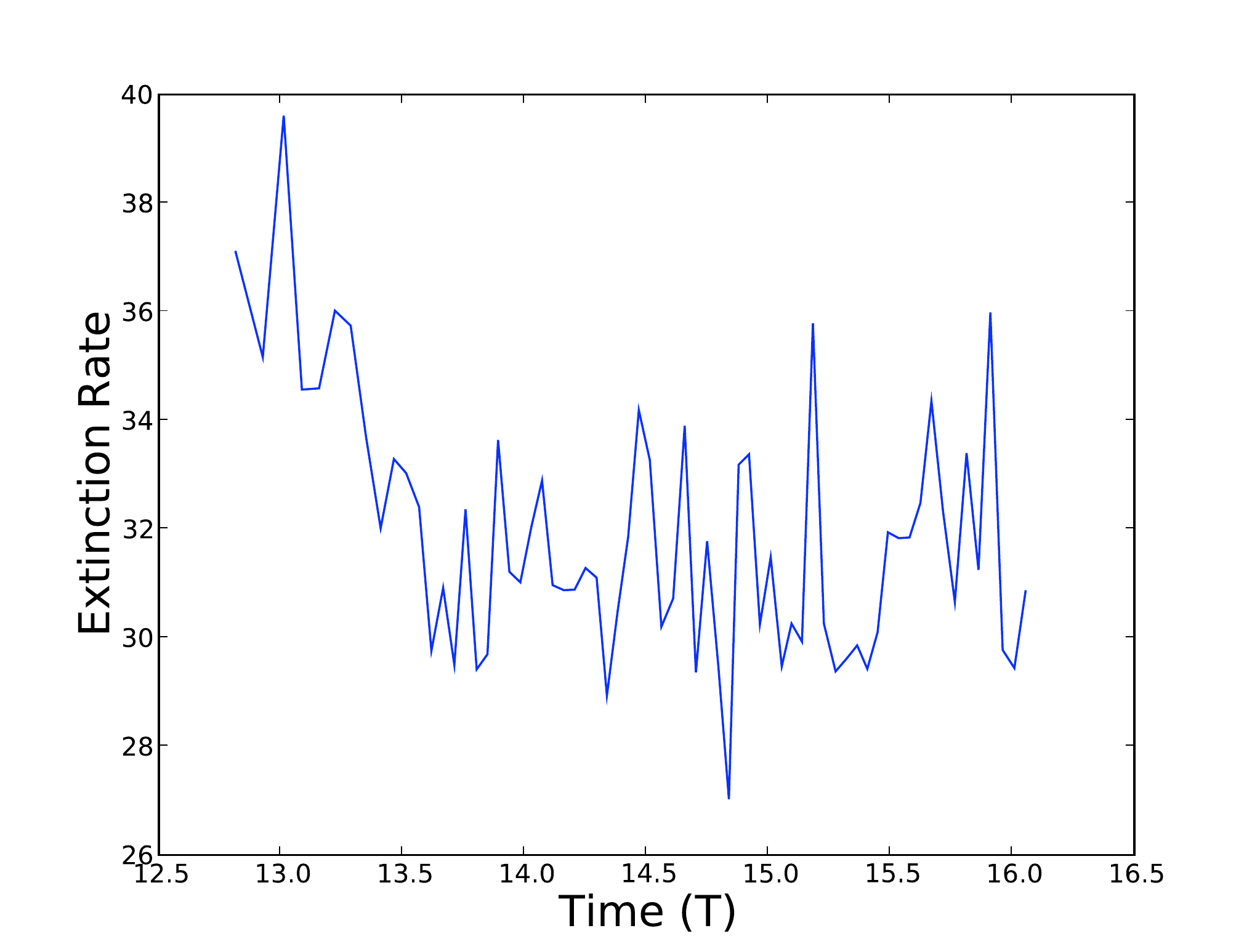, width=9.cm, angle=0}
  \caption{Extinction rate averaged over periods of length $1000$. Time is in units of $1000$ time steps. Parameters are $n=50$, $p = .02037$ (slightly subcritical) and $q = .0008 = 2/n^2$. Note the high background rate of extinction, i.e. no real periods of stasis.}
  \label{RATES}
  \end{center}
\end{figure}

We now consider the time series of the extinction rate, as observed in simulations that survive to full simulation time. Dividing the simulation time into suitably coarse-grained periods of length $1000$ (to emulate the division of fossil data into stratigraphical periods), we consider the ratio of the number of extant genotypes (occupied sites) that went extinct during that time period to the number of extant genotypes that existed during that time period. This provides an estimate of the probability that a randomly selected, living genotype (or occupied site) will go extinct during the time period. We show a typical plot of this time series for subcritical $p$ (using the same $n=50$ data as in previous plots) in Figure \ref{RATES}. Comparison with typical fossil data, e.g. Figure 2(A) of \cite{EvoSOC} shows strong \emph{qualitative} similarity but a much smaller variation in amplitude. Our simulations typically show somewhere between $25\%$ and $42\%$ extinction, while the fossil data varies from close to $0\%$ up to $50\%$. It seems that the interactions of the model, while capable of maintaining a high background rate of extinction, are unable to generate catastrophic biodiversity loss. In other words, we do not see periods of ``statis", i.e. comparably low extinction rate, as in (for example) the Bak-Sneppen model \cite{EvoSOC}. The high background rate of extinction, however, is unsurprising given that the number of species remains relatively constant.

\begin{figure}
  \begin{center}
  \epsfig{file=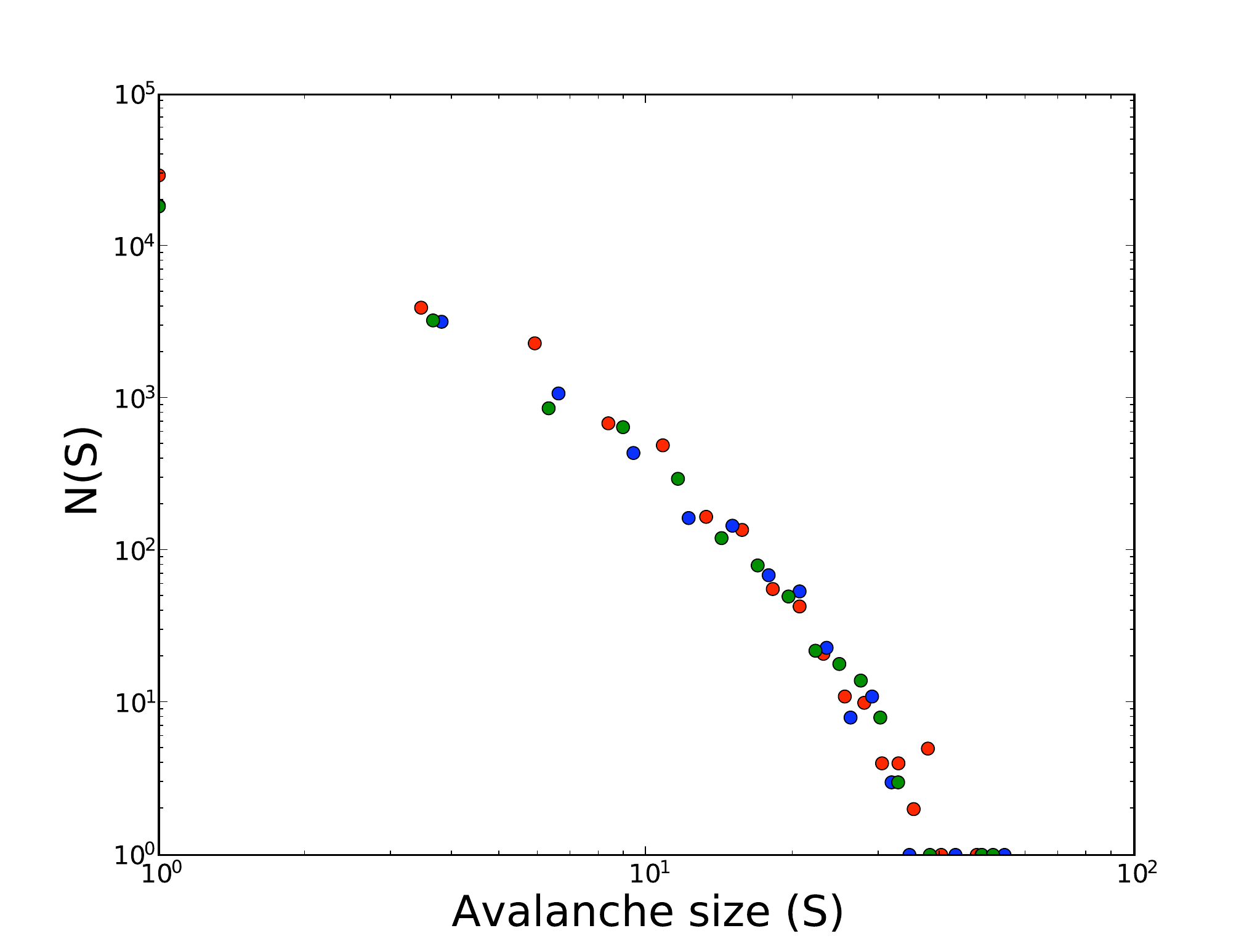, width=9.cm, angle=0}
  \caption{$N(S)$, the number of extinctions of size $S$, for three typical simulations. Extinctions are caused by the occupation of a new site at each time step, and their size is defined as the number of genotypes that go extinct in that time step. Red dots are for a genotype space of dimension $n=30$, blue dots are $n=40$, and green dots are $n=50$. $p$ is slightly subcritical, $p = 1/(n-.9)$ and $q= 2/n^2$ for the appropriate values of $n$. The distribution is fat-tailed but spans too few orders of magnitude for a reliable power-law fit. The distribution is insensitive to changes in $n$ and small changes in $p$ and $q$, including supercritical $p$ (not shown).}

  \label{Avalanche}
  \end{center}
\end{figure}

Recall that when a new site is occupied (i.e. a new extant genotype appears) the interactions thereby created may make other sites inviable, i.e. cause other genotypes to go extinct. If we define the number of sites rendered inviable as $S$, and consider each such instance an ``avalanche", we can measure the size distribution $N(S)$ of extinction events. Several typical distributions are plotted in Figure \ref{Avalanche}. Note that the distributions are essentially identical despite variation in the size of the underlying hypercube ($n=50, n=40, n=30$, simulations used in Figure \ref{DFA}).  Although our data span only one and a half orders of magnitude, they seem to be fat-tailed and one could fit a power law $N(S) \sim S^{-\alpha}$ with $\alpha$ between $2.5$ and $3.0$. Such power law scaling is typical of ecological models \cite{BakSnep, EvoSOC, SoleMan1, Perspect} and while we come as close to the empirical value $\alpha = 2.0$ as Bak-Sneppen type models, other models \cite{SoleMan1, NewmanRob} generate exponents almost identical to the empirical value. Our result is unchanged by any suitable generalization of avalanche causality, and appears insensitive to changes in $n$ as well as small changes in $p$ (including supercritical) and $q$. 

We note the absence of true mass extinctions in our model. This absence is indicated by a value of $\alpha$ larger than the $\alpha$ observed in fossil data, and an average extinction rate that rarely rises above the (high) background extinction rate. The failure to generate mass extinctions may be a result of memory limitations, which forbid the very long runs where one might expect to see large extinctions typical of most self-organized critical models. On the other hand, the absence of a clean power-law or very large events suggests that our model may not be a case of self-organized criticality. 

\begin{figure}
  \begin{center}
  \epsfig{file=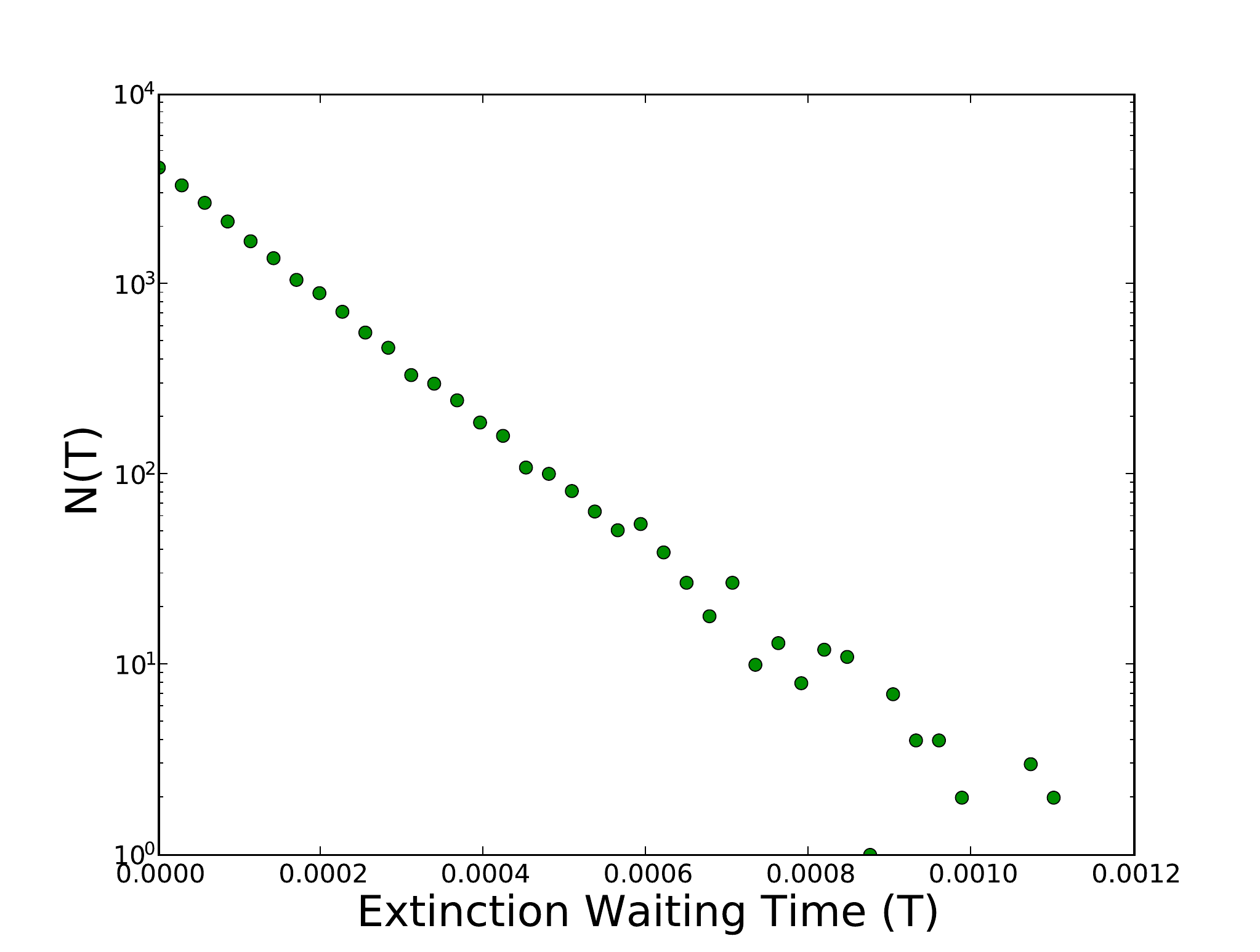, width=9.cm, angle=0}
  \caption{The distribution of waiting times, $T$ between extinction events. The distribution is almost certainly exponential (straight line on a log-linear plot). Parameters are $n=50$, $p = .02037$ (slightly subcritical) and $q = .0008 = 2/n^2$. The exponential distribution is insensitive to changes in $p$ and $q$. Were our model an instance of self-organized criticality, the distribution would be power law \cite{SoleMan2}.}
  \label{Waiting}
  \end{center}
\end{figure}

Further evidence that our model is not an example of self-organized criticality comes from consideration of the waiting time $T$ between extinction events. $T$ is defined as the simulation time that passes between two extinctions events. For self-organized critical models this distribution is typically power law, i.e. $N(T) \sim T^{-\gamma}$ for $\gamma = 3.0 \pm 0.1$ as in \cite{SoleMan2}. A typical distribution for a simulation of our model, shown in Figure \ref{Waiting}, is by contrast exponential. This feature appears robust against small variations in $p$ and $q$, and remains when $T$ is measured in simulation steps. We have some evidence of $n$ insensitivity as well: similar results hold for $n=20$. We should also note that the interpretation of extinctions as a self-organized critical phenomenon has been questioned \cite{Kirchner}. 

\section{Discussion, Interpretation, and Future Work}
\subsection{Discussion}
Our model encapsulates the basic stylized facts about evolution encoded in the fitness landscape and ecological approaches. Each of these approaches provides an important perspective on the nature of selection and the drivers of macroevolution. In a sense, our model remedies important defects in each approach.  To ecological models it adds the notion of an underlying genotype space, whose topology, induced by the fitness landscape, guides the speciation driven by the ecological dynamics. To fitness landscape models it adds explicit notions of coevolutionary dynamics and the construction of niches by other species. In fact, our model illustrates that one need not impose percolation of a neutral network by hand, setting a supercritical $p$ as in \cite{GavRev, Gav1, Gav2}. Rather, ecological interactions create a context in which bridges can be built between the neutral networks of subcritical percolation. As a first step towards constructing a coherent model of evolutionary dynamics --- one that honors most underlying qualitative features without sacrificing simplicity --- our model is an important contribution and starting place.

Turning to the results, we note that our exploration of parameter space has been somewhat limited, concentrating on near critical $p$ and values of $q$ for which escape probability first becomes non-zero. Nevertheless, we find a satisfying robustness of many statistical features to variation of these parameters, as well as the dimension of $\cal{G}$. Furthermore, we rather unexpectedly observed a movement of extant genotypes through genotype space (the ``advancing front") in striking similarity to the empirical results for blastozoans \cite{GavMorph}. This feature is even more surprising when we consider that the model generating similar behavior in \cite{GavMorph} not only omits the fitness landscape over genotype space but also uses exogenous events rather than endogenous dynamics to drive extinctions. Further exploration of this similarity is a major goal of further research. 

Our model also self-organizes a system of positive, self-supporting interactions, stabilizing the number of extant genotypes (occupied sites). In other words, we observed the emergence of a sort of ecology, where the various extant genotypes create and sustain niches for one another. However, we need to explore the correlations between overall interaction bias and extinction to understand this niche construction mechanism more fully.

The time series of occupied sites observed in our simulations displays self-similarity consistent with Brownian motion. At the moment it is rather unclear how to interpret this result. An obvious first step would be to examine e.g. the blastozoan data \cite{GavMorph} and subject it to detrended fluctuation analysis. The robustness of the Brownian motion result indicates that this is probably an inevitable consequence of the current model, so supplying paleontological or empirical evidence of such behavior elsewhere is quite important.

We finally note that our model produced some but not all features of self-organized criticality models of evolution \cite{EvoSOC, SoleMan1, Perspect, NewmanRob}. While we generate time series of extinction rates that are superficially similar to empirical time series, they display limited range and very high background extinction rates. We find fat-tailed and possibly scale free distributions of extinction size, but do not find the very large events characteristic of the best ecological models or the fossil data. The absence of these events may result from current limitations on simulation run time (which can be overcome in the future) but it may also reflect our modeling choices, like the use of an interaction network with Erd\"{o}s-R\'{e}nyi distributed degrees rather than a network with a broad degree distribution. Most crucially, we do not find power law distributed waiting times, which suggests that our model is not an elaborate form of self-organized criticality. This result in itself is quite interesting, as one might wonder how many empirical features of macroevolution can be reproduced \emph{without} a self-organized critical model (although again see \cite{Kirchner}).

In the next subsection we turn to possible interpretations of the model. While we have followed a concrete, ecological interpretation throughout the exposition, we believe the rich set of possible interpretations illustrates the power of our approach. 

\subsection{Interpretation}
The model proposed in this paper is highly abstract. While this abstraction complicates direct comparison with data, it facilitates a wide range of interpretation and hence application. The interpretation guiding the development and exposition of the model was explicitly macroevolutionary and ecological. The underlying sequence space $\cal{G}$ in this case is interpreted morphologically \cite{GavMorph}, and each site can reasonably be interpreted as a new ``extant genotype" (this could represent a single genotype, with clusters representing species, or it could represent a whole species, in which case the clusters would be a higher taxonomical unit). The fitness function measures the evolutionary viability of a particular collection of morphological traits, i.e. the traits must be achievable developmentally and they must fit into an ecological niche. The interactions further shape these ecological niches. As an example, consider an expensive morphological innovation like a beak that is reinforced with an energetically costly, but highly effective structural molecule, which a predatory bird might use for eating a heavily armored prey animal. In the absence of prey requiring this special beak, these mutants would be at an extreme disadvantage and disappear: the cost of the beak is not worth its benefit. But if a shelled creature were to emerge, invulnerable to extant predators --- but vulnerable to the shell-cracking beak of this mutant --- then strong selective pressure would favor the appearance of such predators. The ``strong-beak" mutation might in turn open up a new area of sequence space (i.e. a new cluster) to these birds, providing developmental raw material which could be shaped by mutation into a variety of specialized morphologies. 

To illustrate the flexibility of this model, we provide three alternate interpretations: one in terms of protein evolution, one in terms of RNA evolution, and one in terms of technological evolution. This in no way exhausts the space of interpretations; one can also frame the model as a sort of autocatalytic set of chemical reactions, as a self-assembly process, etc. 

\subsubsection{Protein Evolution}

If we interpret the model as a description of protein evolution, each site in $\cal{G}$ represents a protein sequence of a given length $n$ (the coordination number of the hypercube). The genotype space (better, sequence space) contains all possible protein sequences of that length. In full generality this would be $20^n$ sites, as there are $20$ amino acids. As we set the linear dimension of the hypercube at $2$ (a binary hypercube) we implicitly assume two possible amino acids (or a coarse-graining into hydrophobic and hydrophilic amino acids). A cluster of \emph{a priori} viable sites is a web of neutral neighbors --- sequences that differ, but that share the same basic three-dimensional structure or fold. We say that these sequences comprise a ``neutral network" because they share the same three-dimensional structure, and it is thought that this is what primarily determines the function of a protein, as opposed to the specific sequence. Three-dimensional protein structures are generally far more conserved through evolution than amino acid sequences \cite{Chothia}, and many sequences with no detectable sequence similarity share essentially the same structure, and presumably, function \cite{LoConte}. 

In the current version of the model, we only allow forces internal to the model to change the viability of sites. The ``universe" only includes proteins of a given chain length, so changes to viability occur only through other proteins of length $n$. We do not consider changes to the viability of sites that might occur through forces not represented in the genotype space --- proteins of different lengths, environmental changes, etc.

Chaperone proteins offer one possible mechanism though which we could have "ecological" interactions influencing site viability. Chaperones prevent protein aggregation by binding polypeptide folding intermediates as soon as they emerge from the ribosomal exit tunnel, thus playing a crucial role in the creation of functional proteins with well-defined, three-dimensional conformations \cite{Young}.   Such intermolecular interactions are strongly favored in a crowded cellular environment where numerous unfolded polypeptides are translated in close proximity to one another. Recall that conditionally viable sites can be occupied if sites with + arrows pointing into the conditionally viable site are also occupied. So viable sites could be those proteins of length $n$ that can fold without the help of chaperone proteins into a stable minimum free energy state, and conditionally viable sites could be those sites which are ``activated" by the evolution of a chaperone protein (of length $n$) that allows a potentially unfoldable protein (of length $n$) to fold and hence become viable.  

One way protein sites could become conditionally inactive is when a protein is adversely affected by mutant copies acting in a dominant negative manner. This kind of mutant protein interferes with the activity of an otherwise functional, normal copy of the protein--- for example, via competitive inhibition (where a non-functional protein can still bind a target, blocking the normal protein from binding) or dimerization (where non-functional copies of the protein combine with functional copies), or via the template replication characteristic of prion proteins, which can convert normal proteins to the diseased, misfolded prion phenotype. 

An obvious extension of the model would allow node viability to be changed by forces that are external to the explicitly modeled ``genotype" space. Anything affecting protein viability that is not another protein of length $n$ would count as ÒexternalÓ --- e.g. developmental and environmental changes that open up new realms of possibility for protein evolution. 

It would also be interesting to allow two sites to recombine, thus permitting a jump across sequence space. This extension is especially important in the context of protein evolution because innovation via recombination seems to be one of the primary means by which proteins explore new structures (i.e. new clusters of viable sites) \cite{Patthy, Xia, Bogarad, Cui, Born}. 

\subsubsection{RNA Evolution}
Interpreting our model in terms of RNA evolution is even more straightforward than the protein case. In the protein case we suppress the actual genetic code underwriting the amino acid sequences , dealing only with the space of all amino acid sequences (although it is the genetic code that undergoes mutation). In the RNA case, the genotype-phenotype map is quite direct \cite{TopPos} --- the RNA sequence itself both undergoes mutation and folds into the structures that are acted upon by evolution.

In reality RNA sequences are specified over a four element alphabet $\{ A,U,G,C \}$ so that the genotype space for sequences of length $n$ would contain $4^n$ sites. There are several possible levels of resolution at which RNA structure can be specified; the best compromise between theoretical tractability and empirical accessibility is at the level of secondary structure \cite{Fontana}. So in our model viable sites represent RNA sequences that fold reliably into some ``fit" secondary structure and clusters are neutral networks of sequences all folding into the same secondary structure. It is known from empirical, numerical, and mathematical work that the neutral networks corresponding to any pair of different secondary structures almost touch, and indeed can be found within a few point mutations of an arbitrarily chosen point of $\cal{G}$ --- the so called shape space covering conjecture \cite{TopPos, Reidys}. Thus the picture from our model of neutral networks nearly percolating sequence space is quite realistic in the RNA case.

RNA-RNA interaction is by now a well-known phenomenon, underwriting RNA interference, microRNA-messengerRNA binding, antisense interactions, etc. Interaction is a crucial feature of our model, and $-$ interactions could be interpreted as instances of RNA interference or competitive binding, while $+$ interactions could represent RNA chaperones, of which there are empirical examples \cite{Chaperone}. RNA interaction can be modeled directly with contemporary thermodynamic and folding algorithms \cite{Muckstein, Bernhart}. As in the protein case, our limiting of the mutation operator to point mutations, (excluding insertions, deletions, duplications, and recombinations) is a substantial limitation and simplification. 

\subsection{Technological Evolution}
Since the early work of Schumpeter \cite{Schump} and von Hayek \cite{Hayek}, the tradition of evolutionary economics \cite{Nelson, McKinsey} has emphasized close similarities between evolutionary biology and economic development. Indeed, our model could easily be interpreted in a technological/economic framework. In this case, "genotypes" summarize the collection of traits possessed by a technology or good. Viable sites are economically viable goods (or functioning technologies), and neutral networks are categories of related technology. Inviable sites are products that either don't work or have no extant economic niche. Speciation to a neighboring viable genotype represents an incremental innovation in which one of the traits is tinkered with to generate a new but related good. Interaction represents the well-known fact that goods and technologies can make other goods and technologies economically viable or technologically functional. For example, electronic books required portable electronic readers to take off as a product category. 

\subsection{Future Work}
While investigation of ecological models of macroevolution has slowed considerably, interest in holey fitness landscapes remains high. Indeed, the framework provided by our extension of the holey fitness landscape idea may be instrumental in understanding recent work on epistasis and evolvability \cite{Rouzic}. We thus outline several promising avenues for future work. These divide broadly into work providing better understanding of the present model and work extending the model in more realistic directions. 

Most immediately, a more thorough characterization of the parameter space of the present model is necessary, exploring the $p$ and $q$ space as well as varying the ratio of $+$ to $-$ interactions and relaxing the Erd\"{o}s-R\'{e}nyi assumption on the interaction network (most obviously to allow broad degree distributions). This is an area where analytic work would be of great help, as it would be free of the memory limitations that hindered our numerical work. Thus far our exploration of the model has been exclusively numerical, despite the success of analytic work on holey fitness landscapes \cite{GavRev, Gav1, Gav2, Reidys}.  

Even more interesting is the possibility of empirically testing predictions of the model, for example the comparison of the time-dependent geometry of the set of occupied sites (interpreting them as morphologically distinct species) with fossil data \cite{GavMorph}. Another fruitful line of investigation would involve the definition of plausible phylogenetic trees based on simulations of the model. Many statistical features of empirical phylogenetic trees cannot be reproduced by simple models and perhaps the structuring of genotype space by fitness landscapes and coevolutionary interaction is precisely the missing ingredient \cite{Phylog}. And comparison with in vitro RNA coevolution models would provide an extraordinary opportunity to put the model to direct test. From a more mathematical perspective, modern computational homology techniques could provide considerable insight into the geometry of holey fitness landscapes and their descendants in our model, rare examples of complicated high dimensional spaces with immediate empirical relevance \cite{GavRev}. 

Turning to realistic extensions of the model, an obvious first step would be adding external driving in the form of exogenous extinction. One could also imagine extending the set of mutation operators so that $n$ could change over the course of a simulation run. Thus we could incorporate insertions, deletions, and duplications, as well as recombinations. This would enhance the realism of mutation in the model considerably, at the cost of numerical complication. Another realistic modification would be to relax our assumptions of there being no correlation structure to the fitness landscape, by copying the interaction structure of a mutant largely from the parent, with some small variation. Indeed inheritance of interaction is crucial in ecological models for the self-organization of large extinction events \cite{SoleMan1}, and may be another mechanism to incorporate large extinctions in our model without assuming fat-tailed interaction networks. 

The ease with which we can identify directions for further research and plausible extensions of the model illustrates the still largely untapped potential of unifying the two major metaphors in evolutionary modeling: fitness landscapes and ecology. We believe our model provides an important first step into this as yet unexplored terrain, and we look forward to much fruitful work following in these tentative footsteps. 

\textbf{Acknowledgements:} This work was partially supported by the Santa Fe Institute whose research and education programs are supported by core funding from the National Science Foundation and by gifts and grants from individuals, corporations, other foundations, and members of the Institute's Business Network for Complex Systems Research. Antony Millner made critical contributions to the design of the model and to an earlier version of this paper. The authors gratefully acknowledge stimulating conversations with Sarah Cobey, Devin Drown, Christoph Flamm, Walter Fontana, and D. Eric Smith. JGF thanks iCORE for its generous support; he is currently funded by NSF SciSIP-0915730. MMR is a HHMI postdoctoral associate at the University of Michigan. T.G is funded by a fellowship of the Austrian genome research program GEN-AU and the GEN-AU project "Bioinformatics Integration Network III".


\begin{thebibliography}{99}
\bibitem{GavRev} S. Gavrilets, Evolution and Speciation in a Hyperspace: The Roles of Neutrality, Selection, Mutation, and Random Drift. In J.P. Crutchfield and P. Schuster, eds., {\it Evolutionary Dynamics: Exploring the Interplay of Selection, Accident, Neutrality, and Function} (Oxford University Press, Oxford, 2003). 
\bibitem{Wright}  S. Wright, The roles of mutation, inbreeding, crossbreeding and selection in evolution. In D.F. Jones, ed.,  {\it Proceedings of
the Sixth International Congress on Genetics} {\bf 1}, 356-366 (1932). 
\bibitem{Sigmund} J. Hofbauer and K. Sigmund, {\it Evolutionary Games and Population Dynamics} (Cambridge University Press, 1998). 
\bibitem{EvoSOC} K. Sneppen, P. Bak, H. Flyvbjerg, and M.H. Jensen, Evolution as a self-organized critical phenomenon. {\it Proc. Natl. Acad. Sci. USA} {\bf 92}, 5209-5213 (1995).
\bibitem{SoleMan1} R.V. Sol\'{e}, J. Bascompte, and S.C. Manrubia, Extinction: Bad Genes or Weak Chaos?  {\it Proc. R. Soc. Lond.} B {\bf 263}, No. 1375 (1996). 
\bibitem{Gould} S.J. Gould, {\it Wonderful Life} (W.W. Norton and Co., 1989)
\bibitem{TopPos} B. M. R. Stadler, P.F. Stadler, G.P. Wagner, and W. Fontana, The Topology of the Possible: Formal Spaces Underlying Patterns of Evolutionary Change, {\it J. theor. Biol.} {\bf 213}, 241-274 (2001). 
\bibitem{GavEnc} S. Gavrilets, Evolutionary Ecology: Fitness Landscapes. In S.E. Jorgensen and B. Fath, eds., {\it Encyclopedia of Ecology} (Elsevier, 2008). 
\bibitem{Pigliucci} M. Pigliucci,
Genotype-phenotype mapping and the end of the 'genes as blueprint' metaphor
{\t Phil. Trans. R. Soc. B} 2010 {\bf 365}, 557-566
\bibitem{Pig2} M. Pigliucci, J. Kaplan,
{\it Making sense of evolution} (University of Chicago Press, Chicago, 2006).
\bibitem{Kimura} M. Kimura, {\it The neutral theory of molecular evolution} (Cambridge University Press, New York, 1983).
\bibitem{JMS} J. Maynard Smith, Natural Selection and the Concept of a Protein Space. {\it Nature} {\bf 255}, 563-564 (1970).  
\bibitem{Gav1} S. Gavrilets and J. Gravner, Percolation on the Fitness Hypercube and the Evolution of Reproductive Isolation. {\it J. theor. Biol.} {\bf 184}, 51-64 (1997).
\bibitem{Gav2} J. Gravner, D. Pitman, S. Gavrilets, Percolation on fitness landscapes: Effects of correlation, phenotype, and incompatibilities. {\it J. theor. Biol.} {\bf 248}, 627-645 (2007). 
\bibitem{BakSnep} P. Bak and K. Sneppen, Punctuated equilibrium and criticality in a simple model of evolution. {\it Phys. Rev. Lett.} {\bf 71}, 4087-4090 (1993). 
\bibitem{SoleMan2} R.V. Sol\'{e} and S.C. Manrubia, Extinction and self-organized criticality in a model of large-scale evolution. {\it Phys. Rev} E {\bf 42}, 1 (1996).
\bibitem{Jensen} H.J. Jensen (2004) Emergence of species and punctuated equilibrium in the Tangled Nature model of biological evolution {\it Physcia} {\bf A 340}, 697-704
\bibitem{Christensen} K. Christensen, S. Avogadro di Collobiano, M. Hall, H.J. Jensen (2002) Tangled Nature: A Model of Evolutionary Ecology. {\it J. theor. Biol.} {\bf 216}, 73-84
\bibitem{Hall} M. Hall, K. Christensen, S. Avogadro di Collobiano, H.J. Jensen (2002) Time-dependent extinction rate and species abundance in a tangled-nature model of biological evolution. {\it Physical Review E} {\bf 66}, 011904
\bibitem{Avogadro} S. Avogadro di Collobiano, K. Christensen, H.J. Jensen (2003) The tangled nature model as an evolving quasi-species model. {\it Journal of Physics A: Mathemetical and General} {\bf 36}, 883-891
\bibitem{Tejero} H. Tejero, A. Marin, F. Montero (2010) Effect of lethality on the extinction and on the error threshold of quasispecies. {\it Journal of Theoretical Biology} {\bf 262}, 733-741
\bibitem{Takeuchi} N. Takeuchi, P. Hogeweg (2007) Error-threshold exists in fitness landscapes with lethal mutants. {\it BMC Evolutionary Biology} {\bf 7}, 15
\bibitem{Wilke} C.O. Wilke (2005) Quasispecies theory in the context of population genetics. {\it BMC Evolutionary Biology} {\bf 5}, 44
\bibitem{MayaMan} S.C. Manrubia, M. Paczuski, A Simple Model of Large Scale Organization in Evolution. {\it Int. J. of Mod. Phys.} C {\bf 9}, 1025-1032 (1998).
\bibitem{NewmanRob} B.W. Roberts and M.E.J. Newman, A model for evolution and extinction. {\it J. theor. Biol.} {\bf 180}, 39-54 (1996). 
\bibitem{PerMaya} P. Bak and M. Paczuski, Complexity, contingency, and criticality. {\it Proc. Natl. Acad. Sci. USA} {\bf 92}, 6689-6696 (1995). 
\bibitem{Perspect} R.V. Sol\'{e}, S.C. Manrubia, M. Benton, S. Kauffman, and P. Bak, Criticality and scaling in evolutionary ecology. {\it Trends in Evolution and Ecology} {\bf 14}, 4 (1999). 
\bibitem{Eigen} M. Eigen, R. Winkler-Oswatitsch, A. Dress, Statistical geometry in sequence space: a method of quantitative comparative sequence analysis. {\it Proc. Natal. Acad. Sci. USA} {\bf 85}, 5913-5917 (1988). 
\bibitem{Molon} K. Christensen, N.R. Moloney, {\it Complexity and Criticality} (Imperial College Press, London, 2005). 
\bibitem{PercHyp} B. Kahng, Percolation in the hypercube and the Ising spin-glass relaxation. {\it Phys. Rev.} A {\bf 43}, 1791-1801 (1991).  
\bibitem{Fontana} W. Fontana, Modelling `evo-devo' with RNA. {\it Bioessays} {\bf 24}, 1164-1177 (2002). 
\bibitem{Babajide} A. Babajide, R. Farber, I.L. Hofacker, J. Inman, A.S. Lapedes, and P.F. Stadler, Exploring Protein Sequence Space Using Knowledge-based Potentials, {J. theor. Biol.} {\bf 212} 35-46 (2001). 
\bibitem{Gilles} D.T. Gillespie, Exact stochastic simulation of coupled chemical reactions. {\it J. Phys. Chem.} {\bf 81}(25), 2340-2361 (1977).
\bibitem{GavMorph} S. Gavrilets, Dynamics of clade diversification on the morphological hypercube. {\it Proc. R. Soc. Lond.} B {\bf 266}, 817-824 (1999). 
\bibitem{DFA}  C.K. Peng, S.V. Buldyrev, S. Havlin, M. Simons, H.E. Stanley, A.L. Goldberger, Mosaic organization of DNA nucleotides. {\it Phys. Rev.} E {\bf 49}, 1685-1689 (1994). 
\bibitem{Kirchner} J.W. Kirchner and A. Weil, Correlations in fossil extinction and origination rates through geological time. {\it Proc. Roy. Soc. Lond.} B {\bf 267}, 1301-1309 (2000).
\bibitem{Chothia} C. Chothia and A.M. Lesk, The relation between the divergence of sequence and structure in proteins. {\it EMBO J.} {\bf 5}, 823-826 (1986).
\bibitem{LoConte} L. Lo Cone, B. Ailey, T.J. Hubbard, S.E. Brenner, A.G. Murzin and C. Chothia, SCOP: A structural classification of proteins database. {\it Nucleic Acids Research} {\bf 28}, 257-259 (2000). 
\bibitem{Young} J.C. Young, V.R. Agashe, K. Siegers, and F.U. Hartl, Pathways of chaperone-mediated protein folding in cytosol. {\it Nature Reviews Molecular Cell Biology} {\bf 5}, 781-791 (2004). \
\bibitem{Patthy} L. Patthy, Introns and exons. {\it Current Opinion in Structural Biology} {\bf 4}, 383-392 (1994).
\bibitem{Xia} Y. Xia and M. Levitt, Roles of mutation and recombination in the evolution of protein thermodynamics. {\it Proc. Natl. Acad. Sci. USA} {\bf 99}, 10382-10387 (2002). 
\bibitem{Bogarad} L.D. Bogarad and M.W. Deem, A hierarchical approach to protein molecular evolution. {\it Proc. Natl. Acad. Sci. USA} {\bf 96}, 2591-2595 (1999).
\bibitem{Cui} Y. Cui, W.H. Wong, E. Bornberg-Bauer and H.S. Chan, Recombinatoric exploration of novel folded structures: A heteropolymer-based model of protein evolutionary landscapes. {\it Proc. Natl. Acad. Sci. USA} {\bf 99(2)}, 809-814 (2002). 
\bibitem{Reidys} C. Reidys, P. Stadler, P. Schuster, Generic properties of combinatory maps: neutral networks of RNA secondary structures. {\it Bull. Math. Biol.} {\bf 59}, 339-397 (1997).  
\bibitem{Chaperone} T. Geissmann and D. Touati, Hfq, a new chaperoing role: binding to messenger RNA determines access for small RNA regulator. {\it EMBO J.} {\bf 23}, 396-405 (2004).
\bibitem{Muckstein} U. M\"{u}ckstein, H. Tafer, J. Hackerm\"{u}ller, S. Bernhart, P. Stadler and I. Hofacker, Thermodynamics of RNA-RNA binding. {\it Bioinformatics} {\bf 22}, 1177-1182 (2006).
\bibitem{Bernhart} S. Bernhart, H. Tafer, U. M\"{u}ckstein, C. Flamm, P. Stader and I. Hofacker, Partition function and base pairing probabilities of RNA heterodimers. {\it Algorithms Mol. Biol.} {\bf 1}, 3 (2006). 
\bibitem{Schump} J. Schumpeter, {\it The Theory of Economic Development} (Harvard University Press, 1934).
\bibitem{Hayek} F.A. Hayek, {\it The Constitution of Liberty} (University of Chicago Press, 1978). 
\bibitem{Nelson} R.R. Nelson and S.G. Winter, {\it An Evolutionary Theory of Economic Change} (Harvard University Press, 1982).
\bibitem{McKinsey} E.D. Beinhocker, {\it The Origin of Wealth: Evolution, Complexity, and the Radical Reworking of Economics} (Harvard Business School Press, 2006). 
\bibitem{Rouzic} A. Le Rouzic, O. Carlborg, {\it Evolutionary potential of hidden genetic variation}. {\it Trends in Ecology and Evolution)} {\bf 23}, 33-37 (2008). 
\bibitem{Phylog} E.A. Herrada, C.J. Tessone, K. Klemm, V.M. Equ\'{i}luz, E. Hern\'{a}ndez-Garc\'{i}a, and C.M. Duarte, Universal Scaling in the Branching of the Tree of Life. {\it PLoS One}, {\bf 3}, e2757 (2008). 

\bibitem{Born} E. Bornberg-Bauer. Randomness, Structural Uniqueness, Modularity and Neutral Evolution in Sequence Space of Model Proteins. {\t Z. Phys. Chem.} {\bf 216} (2002) 139Ð154

\end{thebibliography}
\end{document}